\documentclass[fleqn,usenatbib]{mnras}

\usepackage{newtxtext,newtxmath}

\usepackage[T1]{fontenc}

\DeclareRobustCommand{\VAN}[3]{#2}
\let\VANthebibliography\thebibliography
\def\thebibliography{\DeclareRobustCommand{\VAN}[3]{##3}\VANthebibliography}

\usepackage{graphicx}	
\usepackage{amsmath}	

\title[Diffuse ionized gas in star forming galaxies]{The persistence of high altitude non-equilibrium diffuse ionized gas in simulations of star forming galaxies}

\author[L. McCallum et al.]{Lewis McCallum,$^{1}$
Kenneth Wood,$^{1}$
Robert Benjamin,$^{2}$
Camilo Peñaloza,$^{1}$
Dhanesh Krishnarao,$^{3}$
\newauthor
Rowan Smith,$^{1}$
Bert Vandenbroucke$^{4}$
\\
$^{1}$ School of Physics and Astronomy, University of St Andrews, North Haugh, St Andrews, KY16 9SS, UK\\
$^{2}$ Department of Physics, University of Wisconsin-Whitewater, Whitewater, WI 53190, USA\\
$^{3}$ Department of Physics, Colorado College, Colorado Springs, CO 80903, USA\\
$^{4}$  Leiden Observatory, Leiden University, PO Box 9513, 2300 RA Leiden, the Netherlands
}

\date{Accepted XXX. Received YYY; in original form ZZZ}

\pubyear{2023}

\begin{document}
\label{firstpage}
\pagerange{\pageref{firstpage}--\pageref{lastpage}}
\maketitle

\begin{abstract}
Widespread, high altitude, diffuse ionized gas with scale heights of around a kiloparsec is observed in the Milky Way and other star forming galaxies. Numerical radiation-magnetohydrodynamic simulations of a supernova-driven turbulent interstellar medium show that gas can be driven to high altitudes above the galactic midplane, but the degree of ionization is often less than inferred from observations. For computational expediency, ionizing radiation from massive stars is often included as a post-processing step assuming ionization equilibrium. We extend our simulations of a Milky Way-like interstellar medium to include the combined effect of supernovae and photoionization feedback from midplane OB stars and a population of hot evolved low mass stars. The diffuse ionized gas has densities below 0.1~${\rm cm^{-3}}$, so recombination timescales can exceed millions of years. Our simulations now follow the time-dependent ionization and recombination of low density gas. The long recombination timescales result in diffuse ionized gas that persists at large altitudes long after the deaths of massive stars that produce the vast majority of the ionized gas. The diffuse ionized gas does not exhibit the large variability inherent in simulations that adopt ionization equilibrium. The vertical distribution of neutral and ionized gas is close to what is observed in the Milky Way. The volume filling factor of ionised gas increases with altitude resulting in the scale height of free electrons being larger than that inferred from H$\alpha$ emission, thus reconciling the observations of ionized gas made in H$\alpha$ and from pulsar dispersion measurements.

\end{abstract}

\begin{keywords}
methods : numerical -- ISM : structure -- galaxies : ISM -- galaxies : star formation -- ISM : kinematics and dynamics -- ISM : HII regions
\end{keywords}



\section{Introduction}

The existence of a diffuse layer of ionized gas in the Milky Way was first proposed by \cite{hoyleellis} to explain features in the radio spectrum towards the galactic pole. In the years following, the presence of this layer was confirmed by the detection of faint H$\alpha$ emission by \cite{reynolds73}. Since then, all sky surveys have mapped the Milky Way in H$\alpha$ and other emission lines (e.g. The Wisconsin H$\alpha$ Mapper; \citet{wham} and Southern H$\alpha$ Sky Survey Atlas; \citet{shassa}), and the properties of the diffuse ionized gas (DIG) have been determined \citep{digreview,gaensler08,hill08}. Emission from DIG in other galaxies has also been identified \citep{dettmar90,othergalaxydig,ferguson96,Jones2017,jo18,levy19}.

The Galactic DIG is inferred to have an average electron density of $0.01-0.1 {\rm cm^{-3}}$, with a scale height of approximately 1~kpc (\cite{digreview}). Scale heights of H$\alpha$ emission (roughly half the electron density scale height) throughout the Milky Way have been measured in the range 250 pc to greater than 1~kpc \citep{hill14,dk17}. \citet{ocker2020} measures the free electron scale height in the local galactic disk to be approximately 1.5~kpc via pulsar dispersion measures. The temperature of the DIG is higher than that of typical H{\sc ii} regions, at 6000K to 10,000K \citep{tempshigher}.
The DIG requires both a mechanism for supporting the ionized gas at high altitudes, and also for providing the energy to ionize the gas. Many groups have further developed the seminal interstellar medium (ISM) paradigms \citep{field69,cox74,mckee78} and presented results of ever more complex numerical simulations of feedback processes in the ISM, with the energy and momentum injection from supernovae being the major contributor to driving turbulence and large scale galactic outflows \citep{dealv05,joung09,hill12,silcc,kimostriker17,kadofong}. 
Massive O and B stars have the necessary ionizing luminosity budget to produce the Galactic DIG \citep{reynolds90,miller93,dove94}. Until recently, the inclusion of photoionization in the large scale feedback simulations was mostly done as a post-processing step on snapshots of the density grids and adopting ionization equilibrium. Compared to a smooth ISM, the three dimensional structure of a turbulent ISM allows for the transport of ionizing photons from OB stars near the galactic midplane to high altitudes \citep{digreview,wood10,vandenbroucke18}. 

\cite{bertkenny} investigated the effect of photoionization on the support of DIG in simulations excluding supernovae. They found that outflows produced by the warm ionized gas were not enough to produce a DIG layer at the observed densities. 

\cite{kadofong} performed MHD simulations of the ISM including the effects of supernovae, using a self-consistent star formation algorithm based on sink particle formation and the assumption of ionization equilibrium. The vertically resolved neutral structure reproduced that of the Milky Way, but the DIG layer in their simulations was highly variable in time and only occasionally reached the kpc scale heights inferred from observations. The assumption of ionization equilibrium and the absence of accreting gas from the intergalactic medium were suggested as possible explanations for the low density of the DIG in their simulations.
Another source of ionization of the DIG in addition to OB stars is photoionization from a population of hot low mass evolved stars (often referred to as HOLMES). These sources have much lower ionizing luminosities than OB stars, but exist in greater numbers and at higher altitudes above the midplane \citep[e.g.][]{Byler2019}. \cite{randwithmodels}, \cite{floresfjardo} and \cite{bertkenny} showed that some of the trends of emission lines observed in the DIG can be partially explained by the combination of ionizing photons from OB stars plus a more vertically extended population of hot evolved low mass stars. The ionizing luminosities and spectra of these evolved low mass stars are uncertain for the Milky Way; characterisation of these populations with Gaia may provide improved constraints on their contribution to the total ionizing luminosity. 


Our work presented in this paper will focus firstly on the investigation of the two mechanisms required to generate a DIG layer, specifically whether the combined effects of feedback from supernovae and ionizing photons can elevate and maintain the ionization state of gas high above the midplane. Simulations implementing a time-dependent ionization calculation will be compared to those assuming ionization equilibrium. We will also investigate the effect on the density and scale height of the DIG from a component of hot evolved low mass stars contributing to the ionizing photon budget. The work presented here is an extension of \cite{bertkenny} to include supernovae feedback along with improvements in the simulation of star formation, photo- and collisional ionization mechanisms, and time-dependent ionization and recombination.

The structure of this paper is as follows: first the methods for simulating the vertical distribution of the Milky Way are described in section \ref{methods}. This section describes our techniques for calculating photoionization and supernova feedback, the star formation algorithm, ionization state calculations including thermal collisional ionization and time-dependent ionization and recombination. Initial conditions are then discussed. The results are presented in section \ref{results}, showing the dynamical evolution of the simulations, support of high altitude gas, the importance of non-equilibrium ionization, and the effect of hot evolved stars. The results are discussed in section \ref{discussion} and our conclusions are presented in section \ref{conclusions}.

\section{Methods}
\label{methods}
\subsection{Radiation-hydrodynamics simulations}

The results presented in this paper are from simulations using the Monte-Carlo radiation-hydrodynamics code \texttt{CMacIonize} \citep{cmacionize,cmi2} in task-based mode. Starting from initial conditions described below, we run all simulations to a time of 300~Myr, with the fiducial simulation being run to 500~Myr. 

\subsubsection{Grid properties}

\texttt{CMacIonize} was initialised with a static Cartesian grid, with the hydrodynamic equations being solved in an Eulerian framework. The extent of the grid in physical space was set as 1 kpc $\times$ 1 kpc $\times$ 6 kpc, comprised of $128\times128\times768$ grid cells. The boundary conditions were set up to be periodic in the $x$ and $y$ axes, and allow outflow in the $z$ axis. Our grid dimensions and spatial resolution are comparable to those of other studies.

\subsubsection{Equation of state}

In the simulations of \citet{bertkenny}, the effect of photoionization on the extended diffuse ionized gas was explored using a two-state isothermal equation of state, defining the gas temperature and pressure as a function of ionization state determined from the radiation step. In this work however, with thermal energy injections from supernova events, we use a non-isothermal equation of state (polytropic index = 5/3) and keep track of heating and cooling terms explicitly. The heating terms are composed of contributions due to the photoionization heating from the radiation transfer step, and the thermal energy from supernovae. Only terms from hydrogen photoionization heating are included in this work.

Cooling rates are determined from the tables of \citet{derijke} and a cooling algorithm was implemented to work between every hydrodynamical step of the simulation. The algorithm was developed to work independently of the hydrodynamical timestep. Each cell is only permitted to radiate $10^{-3} U$, where $U$ is the total thermal energy of the cell, before the cooling table is again referenced against the new temperature of the cell. This process is repeated until the total hydrodynamical timestep has been integrated through. The pressure is then updated to match the new temperature for each cell's given mass, volume and mean molecular weight. The cooling curve adopted was that for solar metallicity for a gas density of $1 {\rm cm}^{-3}$, and the absolute cooling rate scaled with $n^{2}$ where $n$ is the total hydrogen density.

\subsubsection{Photoionization feedback}

The effect of photoionization from massive stars is included through Monte Carlo radiation transfer (MCRT) simulations every 0.2 Myr of simulation time. Similar to the work of \cite{bertkenny}, this was found to be a sufficient timestep to obtain converged results. Each radiation step is executed as a series of MCRT photoionization simulations. The emission locations for ionizing photon packets are sampled from a set of source locations. The frequency of each individual photon packet is sampled from model stellar atmosphere/stellar population spectra. The spectrum used spans 1000 bins from 13.6~eV to 54~eV. Each photon frequency is used to determine it's individual photoionization cross section from the fits of \citet{verner95}. In the fiducial runs, we use the WMBASIC stellar atmosphere models \citep{wmbasic}, assuming an effective temperature of 40,000~K for all massive stars.

The photon packets are traced until they either exit the simulation box or are absorbed and re-emitted as a non-ionizing packet. Path length counters are employed to calculate the photoionization and heating rates within each cell as described in \cite{woodercolano}. 

With the optical depth across each cell being dependent upon the ionization state and temperature, this process must be done iteratively, with each cell's temperature and ionization state being updated in each iteration until convergence, typically ten iterations. One underlying simplification in this MCRT technique is an infinite speed of light. This is appropriate for the small size of the simulation box, with the longest axis being 6~kpc, which has a $\sim$20,000~year light crossing time, compared to the 200,000~year radiation timestep.

\subsubsection{Time-dependent ionization calculation}
\label{ionstatelimit}

\begin{figure}
    \centering
    \includegraphics[width=\columnwidth]{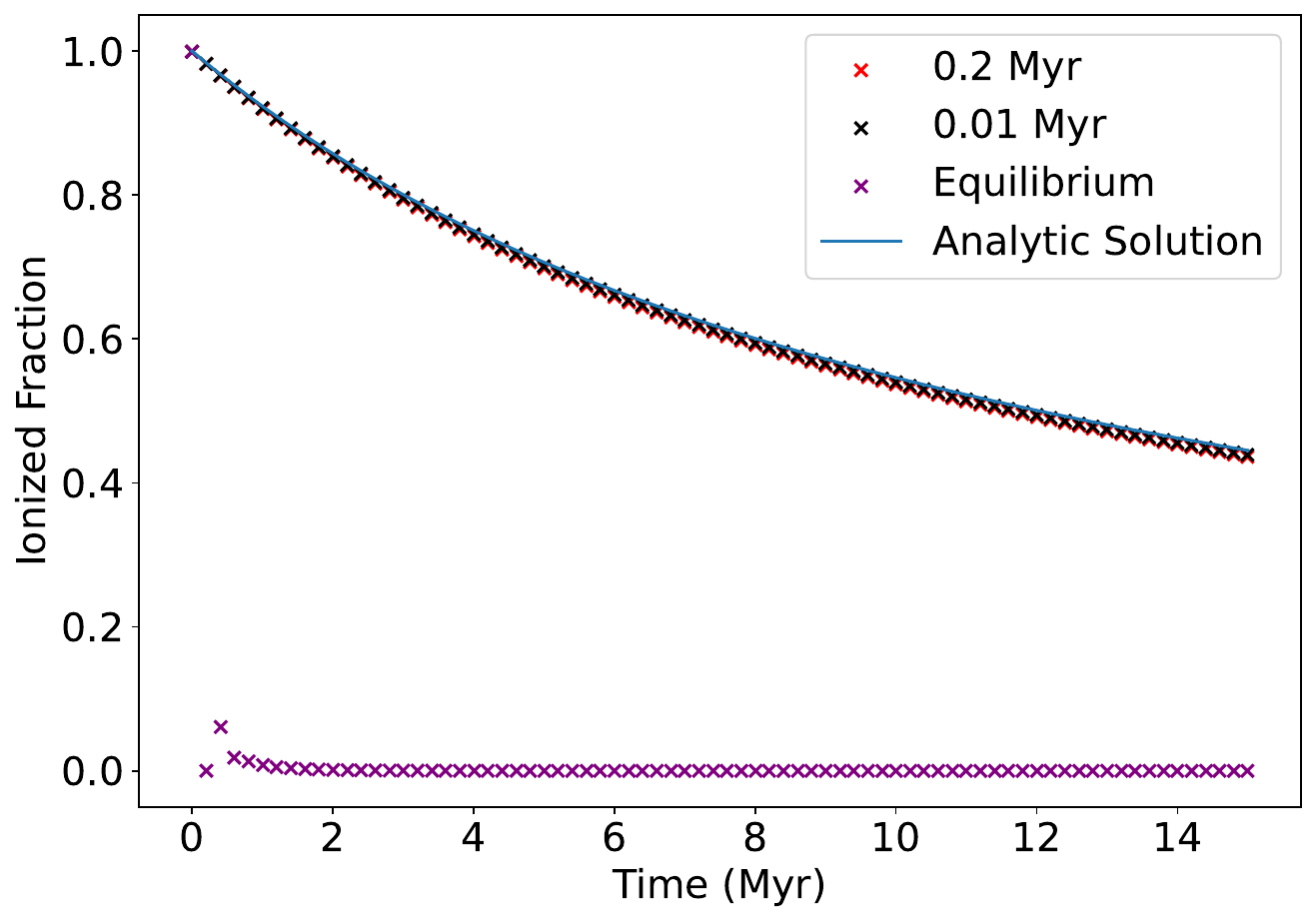}
    \includegraphics[width=\columnwidth]{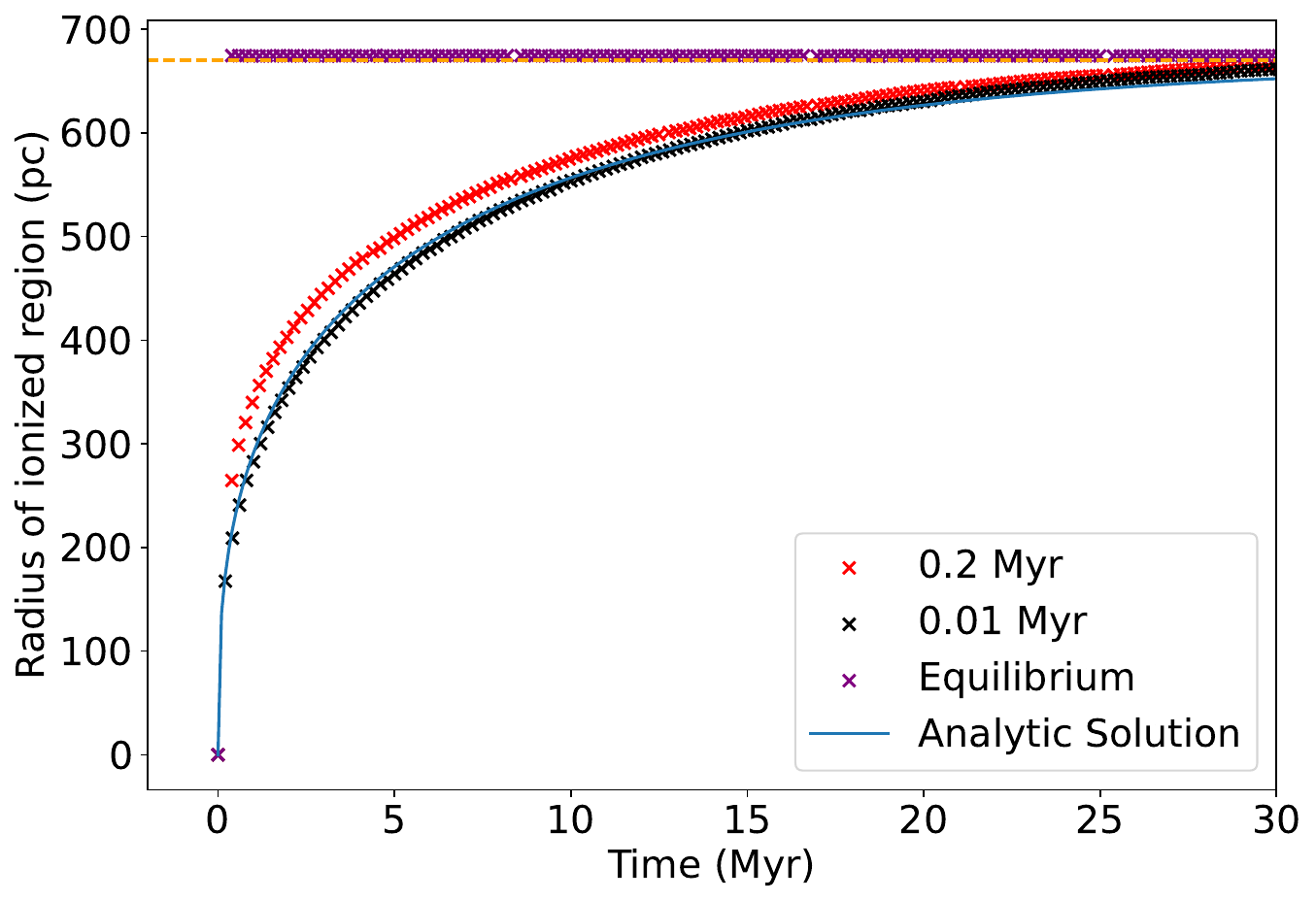}
    \caption{Upper panel shows the test of slowly recombining hydrogen gas at a density of $0.01 {\rm cm}^{-3}$. Lower panel shows development of a Stromgren sphere around a source of ionizing luminosity $Q = 10^{48}\,{\rm s}^{-1}$, also in a uniform hydrogen density of $0.01 {\rm cm}^{-3}$. Each panel shows the analytic solution along with the results from our radiation hydrodynamics code assuming ionization equilibrium and two simulations with our non-equilibrium ionization scheme for two different radiation timesteps of 0.2 and 0.01 Myr. The orange dashed horizontal line shows the expected final Stromgren radius.}
    \label{bench}
\end{figure}

\cite{bertkenny} implemented a Monte Carlo ionization state calculator based on the assumption of ionization equilibrium. In this work we have extended the simulations of \cite{bertkenny} to include supernova feedback and more realistic stellar populations, producing a more stochastic and variable ionizing luminosity. With this variability comes the possibility that much of the gas might not be in ionization equilibrium for long periods of time. At typical DIG densities, $n\sim 0.1{\rm cm}^{-3}$ to $0.01{\rm cm}^{-3}$, the recombination timescale, $t_R = (n\,\alpha)^{-1}$, can exceed ten million years for the low density gas (adopting a hydrogen recombination rate coefficient, $\alpha = 3\times 10^{-13}{\rm cm}^3\,{\rm s}^{-1}$). To account for this we implement a time-dependent ionization calculation scheme. The inclusion of supernova feedback introduces hot gas so we also include collisional ionization in the ionization state calculation.

The ionization state is updated at each radiation timestep. As in \cite{bertkenny}, the ionization equilibrium state is first calculated, but with the inclusion of collisional ionization we use the following equation for the hydrogen neutral fraction in ionization equilibrium:
\begin{equation}
x_{n} = \frac{2\alpha_{H}n_{H}}{J_{H} + (2\alpha_H + \gamma_{H})n_{H} + \sqrt{(J_{H} + \gamma_{H} n_{H})^{2} + 4J_{H}\alpha_{H} n_{H}}}
\end{equation}
where $\alpha_{H}$ and $\gamma_{H}$ are the radiative recombination rate and collisional ionization rate (for the given temperature) respectively. The recombination and collisional ionization rates as a function of temperature are from \citet{benjamin01}. $n_{H}$ is the total hydrogen number density. This is very similar to the equation presented by \cite{kadofong}, except we use $J_{H}$ which is the hydrogen-intersecting mean intensity integral as calculated from path length counters in the MCRT step (using equation 14 from \cite{woodercolano}).

Once the above neutral hydrogen fraction is calculated, we compare the value to that from the previous radiation timestep to give the neutral fraction change adopting ionisation equilibrium. We then calculate the \emph{maximum possible change} in neutral fraction per timestep by balancing total ionizations with total recombinations and dividing by the total number of hydrogen atoms. This gives the following expression:
\begin{equation}
\label{numion}
\frac{\Delta x}{\Delta t} = \alpha n_{H} (1-x_{0})^{2} - x_{0}J_{H} - \gamma n_{H}x_{0}(1-x_{0})
\end{equation}
where $\frac{\Delta x}{\Delta t}$ is the change in neutral fraction per time and $x_{0}$ is the neutral fraction from the previous timestep.

If the magnitude of $\Delta x$ in the equilibrium calculation is found to be greater than the maximum $\Delta x$ from the numerical limit, and the signs match, a `limiter' is employed and the ionization state is only permitted to change by the maximum possible amount as determined from equation \ref{numion}.

This ionization state limiter method ensures that denser gas ($n > 1\,{\rm cm}^{-3}$) remains in ionization equilibrium (as expected over the radiation timesteps of these simulations) but the lower density material can change ionization state over realistic timescales. The benefit of this semi-numerical time dependent solution compared to a fully time dependent scheme is it requires a coarser radiation timestep to accurately represent the ionization state of denser gas, but captures the time dependent photoionization and radiative recombination of low density diffuse ionized gas. As we will demonstrate, the inclusion of non-equilibrium ionization is crucial for maintaining high altitude diffuse ionized gas in our simulations.

While the ionization state throughout the simulation is only updated every radiation timestep, the ionization state limiter is applied at 10 times the temporal resolution of the radiation timestep. This is because the computational expense in calculating the ionization state is significantly less than that of the MCRT photon transport step, but tests at substeps of 100 times temporal resolution were found to be beyond that necessary for convergence of this method.

Two tests were run using this time-dependent method for gas with a density of $0.01\rm{cm}^{-3}$. The first follows the slow recombination of a cube of ionized gas. The second simulates the time dependent development of a Stromgren sphere around an isotropic source with ionizing luminosity $Q=10^{48} {\rm s}^{-1}$. This test density has been chosen because it represents typical DIG densities. The results of the tests are shown in figure \ref{bench}. These tests were run with two different ionization calculation timesteps, and also one run under the assumption of ionization equilibrium.

The analytic solutions shown in figure \ref{bench} assume a static density structure, so for the purposes of these tests the hydrodynamics has been turned off and gas is not able to flow from cell to cell. The recombination rate has also been set at a fixed value of $2.7\times 10^{-13} {\rm cm}^{3} {\rm s}^{-1}$. One difference between the test case and the analytic solution is that the former includes the effect of collisional ionization, however at photoionized temperatures of approximately $10^{4}$K the impact of this term is negligible.
The small difference early in the 0.01 Myr and 0.2 Myr timestep simulations amounts to less than a single grid cell, and we consider the algorithm converged for DIG densities at a radiation timestep of 0.2 Myr.

Time-dependent analytic solutions for recombining gas and the development of a Stromgren sphere are shown in figure \ref{bench}. For the recombining gas the equation is
\begin{equation}
\label{analytic-Recom}
x(t) = -\frac{1}{\alpha n t - 1}
\end{equation}
where $x(t)$ is the hydrogen ionization fraction as a function of time, $\alpha$ is the hydrogen radiative recombination rate coefficient and $n$ is the total hydrogen number density. For the time-dependent development of a Stromgren sphere the analytic solution (e.g., from \citep{spitzerbook}) is
\begin{equation}
\label{analytic-R}
R(t) = R_S\left(1-{\rm e}^{-t/t_r}\right)^{1/3}
\end{equation}
where the Stromgren radius for pure hydrogen gas of density $n$ produced by a star of ionizing luminosity $Q$ is $R_S^3 = 3\,Q/(4\pi n^2 \alpha)$.

The numerical resolution study presented by \citet{deng23} suggests a radiative timestep of $0.1t_{\rm{rec}}$ is required to accurately resolve the ionization feedback from young massive stars. Using this benchmark alongside our adopted radiation timestep of 0.2 Myr, we find that we will reach this threshold for densities $<0.06\,{\rm cm}^{-3}$. For higher densities we will not reach this temporal resolution, but the recombination times are short enough that our method will allow the gas to be in ionization equilibrium.

Figure \ref{bench} shows that an ionization calculation timestep of 0.2 Myr is sufficient to approximate both the development of an ionized region, and its recombination in $0.01 {\rm cm}^{-3}$ gas (comparable to DIG density). In the photoionization equilibrium recombination simulation we see a small spike in the ionization fraction early on in the equilibrium recombination test. This spike is due to the ideal gas equation of state resulting in the temperature doubling as the ionized gas (proton and electron) immediately recombines into atomic hydrogen. This temperature jump causes a small spike in the ionization fraction due to collisional ionization.

To determine the importance of non-equilibrium ionization for the low density DIG, our fiducial model will be run both with and without the ionization state limiter.

\subsubsection{Supernova feedback}

Supernova feedback is implemented as described in \citet{gatto}. This technique implements two different modes of energy injection, the first being a direct injection of thermal energy and the second an injection of momentum. Whether the thermal energy or momentum injection algorithm is called depends on the resolution relative to the expected Sedov-Taylor blast radius. 

For each supernova event, an injection radius ($R_{\rm {inj}}$) is determined as the maximum of the following two values: the radius surrounding the event containing $10^{3}M_{\odot}$ or a radius of four grid cells. This ensures the injection radius is always resolved by at least eight cells in diameter.

To determine if either the thermal or momentum injection regime is to be deployed, we first calculate the radius of the bubble at the end of the Sedov-Taylor phase as presented in \citet{blondin}. The radius is:
\begin{equation}
   R_{\rm{ST}} = 19.1 \left(\frac{E_{\rm{SN}}}{10^{51}}\right)^{\frac{5}{17}}\bar{n}^{-\frac{7}{17}} \rm{ pc}\, 
\end{equation}
where $E_{\rm{SN}}$ is the supernova energy in units of $10^{51}\,\rm{ergs}$ and $\bar{n}$ is the average density in units of $\rm{cm}^{-3}$. The value of $R_{\rm{ST}}$ is compared to that of four grid cells, and if the Sedov-Taylor radius is found to be resolved, thermal energy injection is called. Otherwise, momentum is injected.

In the case of thermal energy injection, the energy is distributed smoothly throughout the injection volume. The injection is done by increasing each cell temperature and pressure appropriately for the individual case of mass, volume and mean molecular weight.

Pressure is
\begin{equation}
    P = \frac{(\gamma-1)U}{V}
\end{equation}
and temperature is
\begin{equation}
    T = \frac{m_{H}\mu P}{k \rho}
\end{equation}
where $\gamma$ is the polytropic index (always 5/3), $U$ is the new total thermal energy of the cell, $V$ is cell volume, $\mu$ is mean molecular mass within the cell, $\rho$ is cell mass density, $m_{H}$ is the hydrogen mass and $k$ is Boltzmann's constant. 

In the case of momentum injection, the increase in momentum required for the equivalent thermal energy injection is \citep{blondin}
\begin{equation}
p_{\rm{ST}} = 2.6\times 10^{5} \left(E_{\rm{SN}}\right)^{\frac{16}{17}}\bar{n}^{-\frac{2}{17}} \, M_{\odot}\, {\rm km\, s}^{-1} 
\end{equation}

This momentum is then injected smoothly within the injection volume by adding the following velocity to the system, pointing radially away from the source of the injection,
\begin{equation}
    v_{\rm{inj}} = \frac{p_{\rm{ST}}}{M_{\rm{inj}}}\, 
\end{equation}
where $M_{\rm{inj}}$ is the total mass (in solar masses) within the injection radius. In the case of multiple momentum injections affecting a single cell, the change in velocity is simply summed over all injections, meaning momentum kicks from two adjacent supernovae can cancel out, leaving some gas cells unaffected. However, we expect multiple concurrent injections to be rare, due to the fine temporal resolution (as required by the hydrodynamics solver) being much shorter than even the shortest stellar lifetimes.

Some benchmark tests in uniform density media were carried out after implementation of the supernova thermal injection algorithm. The time dependent temperature and density structure of supernova benchmarks presented by \cite{novabench} were successfully replicated with our code.

\begin{figure}
    \centering
    \includegraphics[width=\columnwidth]{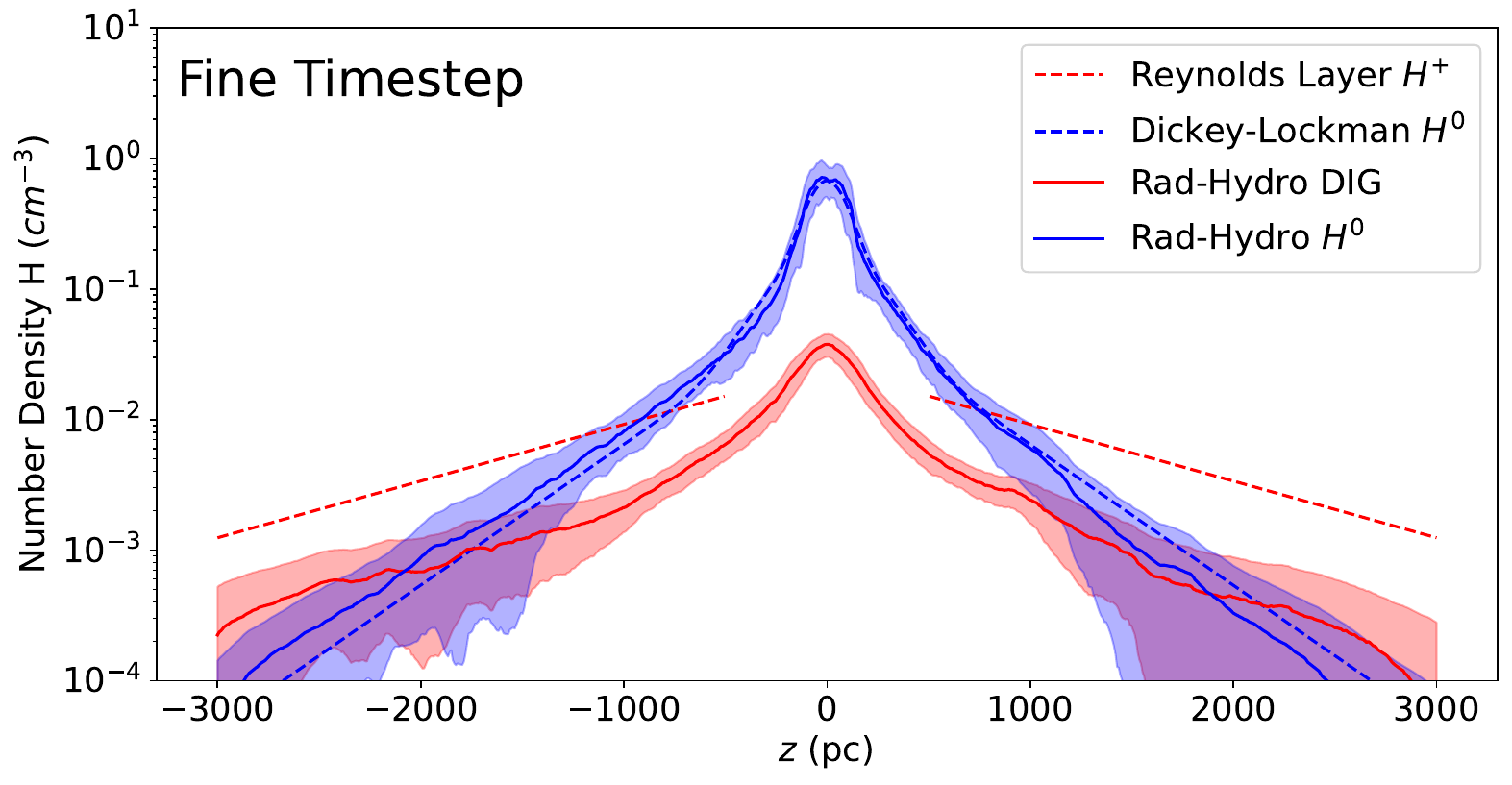}
    \includegraphics[width=\columnwidth]{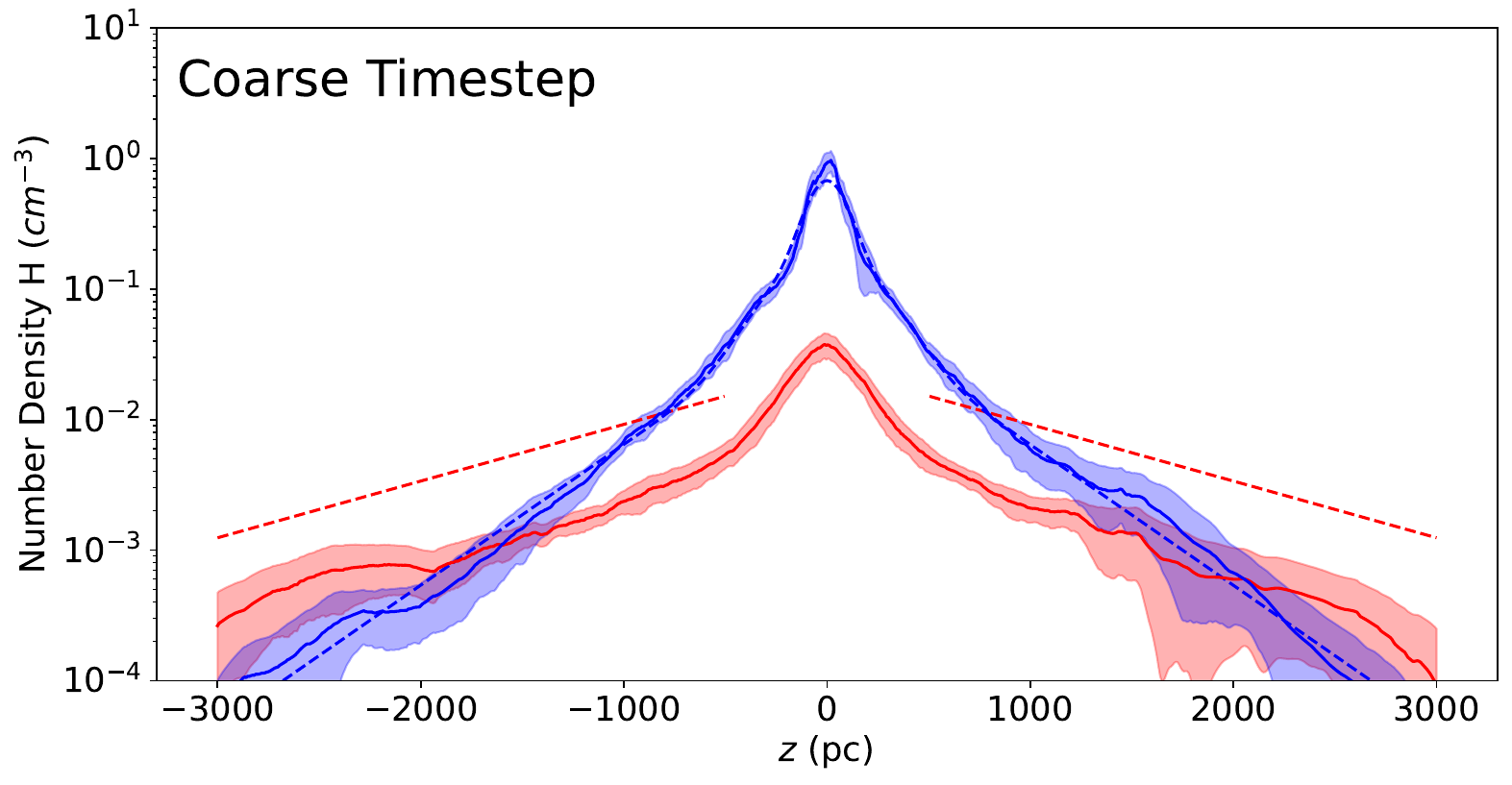}
    \caption{Vertical extent of neutral and warm ionized hydrogen both in the Milky Way and in our coarse and fine radiation timestep simulations. Top panel shows the fine timestep simulation at 0.05 Myr, and the bottom panel shows the coarse timestep run at 0.2 Myr. The dashed blue line shows the Dickey-Lockman approximation for neutral hydrogen in the Milky Way, the dashed red line shows the Reynolds approximation for ionized hydrogen in the Milky Way. The solid red and blue lines show the median results of our simulations between 150-500 Myr of evolution, and the fill shows the $1\sigma$ variance for each line.}
    \label{radconvfig}
\end{figure}

\subsubsection{Star formation}
\label{sfdescribe}

An important ingredient of our simulations is the location and type of stellar sources, both for determining the spectrum and intensity of the interstellar radiation field and the locations of supernova events. Since the prime motivation of this work is to understand the support of the extended diffuse ionized gas, we do not resolve individual molecular clouds, nor do we include the effect of gas self-gravity. With this in mind, an algorithm was developed for generating sources within the highest density structures in our simulations. As in the work of \cite{silcc}, we separate star formation into two categories; \emph{peak driving} and \emph{random driving}. The former simulates the existence of extremely massive stars, which have high ionizing luminosities and explode as supernovae within a short time of their formation in dense molecular clouds. The density structure provides a probability distribution from which to sample the locations of these sources, causing stars to form at the peaks of the density distribution. 

A maximum peak height above the midplane is also set at 200 pc to prevent nonphysical rates of star formation from occurring in denser outflows. \citet{maiolino17} and \citet{gallagher19} present evidence of widespread star formation in outflows, however with our star formation algorithm not being fully self-consistent we find that without a star formation ceiling, the distribution of ionizing radiation becomes dominated by sources formed at high altitudes. This has the effect of more ionized gas being produced at altitude. In future work, our star formation algorithm could be adapted to produce a realistic amount of outflow star formation, and this unexplored parameter space could be investigated. For the purposes of this work, the star formation ceiling is fiducicially set as 200~pc above and below the midplane, three times the scale height of OB stars.

The {\it{random driving}} represents stars of mass greater than $8 M_{\odot}$, but which can live long enough to travel away from the densest regions of the simulation, and inject ionizing photons and energy from supernovae into lower density regions. These sources can also include stars whose molecular cloud of origin has already been disrupted by the ionizing photons or supernovae from nearby massive stars. The source locations for random driving are sampled uniformly in the $x-y$ plane, but from a normal distribution with a scale height of 63~pc in the $z$-axis \citep{63pcref}. The SFR balance between \emph{peak driving} and \emph{random driving} is set to 50/50.

An initial star formation rate for the entire grid is chosen and represents the normalisation factor of the Kennicutt-Schmidt relation between surface density and SFR.
\begin{equation}
    \Sigma_{\rm {SFR}} \propto \Sigma_{\rm {gas}}^{1.4}
    \label{kennicutt}
\end{equation}

As the simulation progresses, the surface density of the disc changes and the SFR is adjusted accordingly. We calculate the surface density from the region of $z\pm {\rm{200pc}}$ and scale the SFR as the surface density to the power of 1.4. The initial SFR is chosen such that a Milky Way surface density will generate a Milky Way SFR surface density, taken to be $~0.0032 M_{\odot} {\rm{yr}}^{-1} {\rm{kpc}}^{-2}$ \citep{mwsfr}.

Throughout the simulation, the current SFR is multiplied by the star forming timestep (0.2 Myr) to give the total mass of stars formed. A factor of 0.073 is applied to this value to determine the total mass of stars to be formed with masses greater than $8 M_{\odot}$. Stars below this mass are not considered as they do not result in supernova events or have significant ionizing luminosity. This factor is determined from the Kroupa initial mass function \citep[IMF;][]{kroupa}.

With the target mass of massive stars known, we sample single stars from the Kroupa IMF between 8 and 120 $M_{\odot}$ until the target mass is reached. Any overstepping of the target mass within any one timestep is saved and subtracted from the target mass at the next star forming step.

For each stellar source we determine the ionizing luminosity from the tables of \cite{pauldrach} as a function of mass. The lifetime of each source is determined by the following relation from \citet{raiteri96} assuming a solar metallicity of 0.012:
\begin{equation}
\log(\tau) = 9.9552 - 3.3370\log(M) + 0.8116(\log(M)^{2})
\end{equation}
where $\tau$ is stellar lifetime in years and $M$ is stellar mass in solar masses.
While the \citet{pauldrach} tables inform a total ionizing luminosity for each individual source, we must acknowledge that much of the ionizing flux from massive stars will be absorbed by the dense gas of their respective birth clouds. With our resolution limited to 7.9 pc, we do not resolve individual molecular clouds, and hence underestimate the maximum densities achieved in the CNM. Similarly, \citet{kadofong} estimates from their simulations that the contribution of dust absorption is primarily occurring in densest material in their models. To account for ionizing flux losses due to unresolved dense material, we introduce a parameter for ionizing photon molecular cloud escape fraction. This value is fiducially set as 0.1, with a $\it{bright}$ and $\it{dim}$ model also run at escape fractions of 0.5 and 0.02 respectively. Throughout this work, the total OB+HOLMES ionizing luminosity will be referred to as the {\it available ionizing luminosity}, while total OB luminosity escaping high density clouds will be called the {\it escaping ionizing luminosity}.

With the star formation algorithm being called once every 0.2 Myr, a uniform random sampled time value between zero and 0.2 Myr is subtracted from each of the source lifetimes to account for the fact that the sources could have formed at any point since the star formation algorithm was last called.

Once a star reaches the end of its lifetime, it is removed from the simulation and a supernova event is initiated at its location. Therefore the ionizing luminosity and supernova rate is self-consistently determined from the initial SFR.

The above method for generating sources of ionizing luminosity and supernovae only accounts for Type II/core collapse supernova events. To include the effect of Type Ia supernovae, a constant rate of $4\,{\rm{Myr}}^{-1} {\rm{kpc}}^{-2}$ and scale height of $325 {\rm{pc}}$ was adopted \citep{type1}, and the supernova events are injected at locations as sampled by the random driving method described above. This is because Type Ia supernovae originate from systems of evolved stars, which need not exist within the dense molecular clouds from which they were born.

The main drawbacks of the method described above for generating a photon and supernova source distribution are 1) the static nature of the sources and 2) the need to choose a normalisation for the SFR/surface density relation by hand. These are two consequences of not calculating the full gravitational potential throughout the simulation, meaning sink particles cannot be introduced to address problem 2), and source accelerations cannot be determined to address problem 1).

\subsubsection{Hot low mass evolved stars}

Also included in the determination of the ionizing luminosity is the existence of low mass hot evolved stars and white dwarfs. These sources, while low in ionizing luminosity, exist in great numbers at high altitudes, with harder spectra than midplane OB stars. The luminosities of these sources are uncertain, so three runs are carried out with differing ionizing luminosities for these sources. This population is included as a set of 1000 sources, with the two runs using an ionizing luminosity per source of $1\times10^{45}{\rm s}^{-1}$ and $5\times10^{45}{\rm s}^{-1}$. The locations of these low mass sources are sampled in the same way as the random driving technique described above with a scale height of 700~pc. The effect of these stars is included in the radiation-hydrodynamics simulation after an evolution time of 150~Myr. This delay is to prevent issues surrounding ionizing photons being injected into very low density gas associated with the high altitude regions before the evolution has driven outflows to these heights. In this very low density gas, the periodic $x-y$ boundaries mean that photons which would be expected to escape laterally can trace very long path lengths, and the infinite plane assumption due to the periodic boundaries becomes non-physical. The 150 Myr delay ensures that there is at least some gas present at height to inhibit these non-physical photon paths.

For the hot low mass stars we use the same 40,000~K spectrum as the population of massive stars. This is likely a softer spectrum than these sources exhibit in reality. The spectral shape of the radiation field is important for accurately simulating emission lines from other elements. Adopting the same spectrum for all sources will have little effect on the ionization state of hydrogen in our simulations which is primarily determined by the ionizing luminosity and not the spectral shape.

\begin{figure*}
    \centering
    \includegraphics[width=\textwidth]{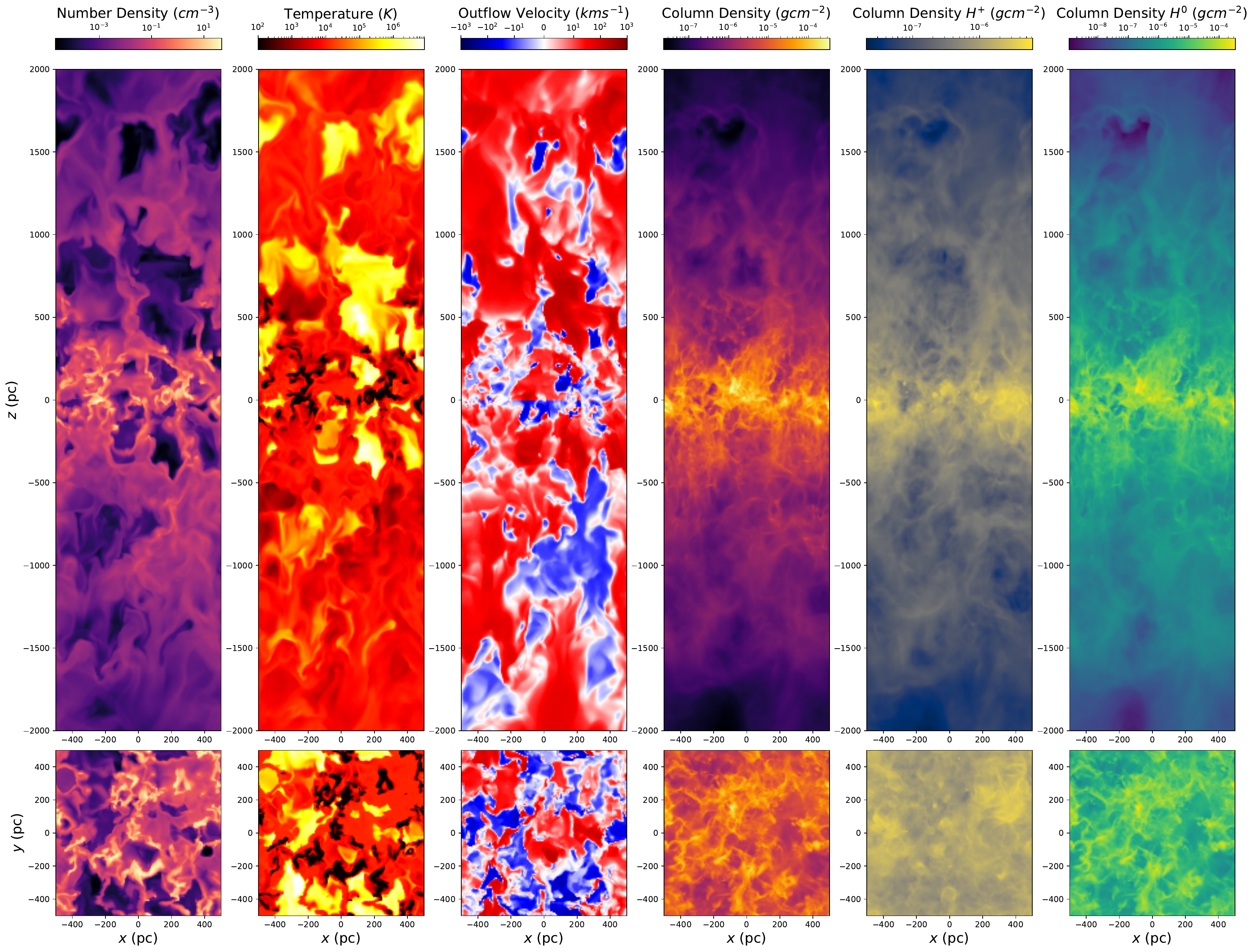}
    \caption{A visualisation of a snapshot from the fiducial model at a time of 350 Myr. First column shows a slice of total hydrogen number density through the centre of the grid, upper panel is a vertical slice and lower panel shows the density in the midplane. Second column shows temperature slices. Third column shows outflow velocity slices, ${d|z|}/{dt}$. Fourth column shows total projected hydrogen column density along the $y$ (upper panel) and $z$ (lower panel) axes. Columns five and six show the projected column density of neutral and ionized gas.}
    \label{bigplot}
\end{figure*}

\begin{figure*}
    \centering
    \includegraphics[width=\textwidth]{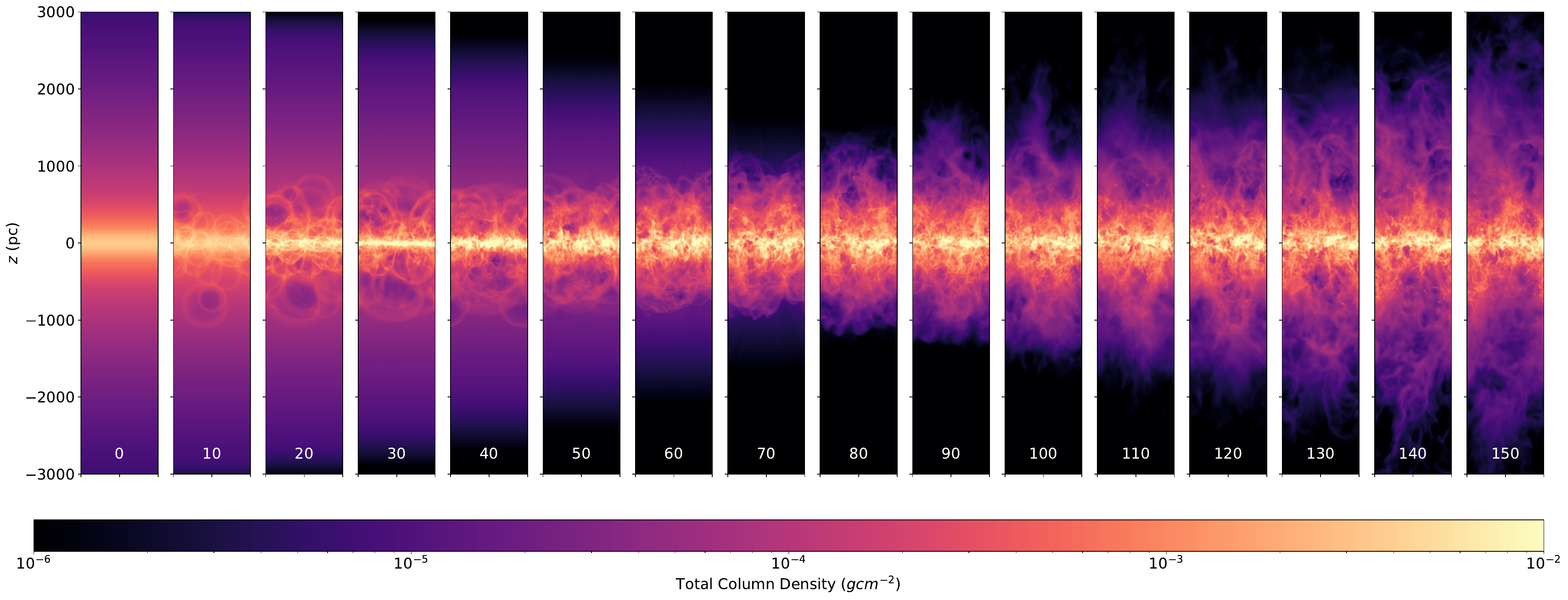}
    \caption{Evolution of edge on projected total column density for the first period of inflow and outflow in our fiducial model. White text at the bottom of each panel shows the evolution time in Myr.}
    \label{panelevo}
\end{figure*}

\begin{figure*}
    \centering
    \includegraphics[width=\textwidth]{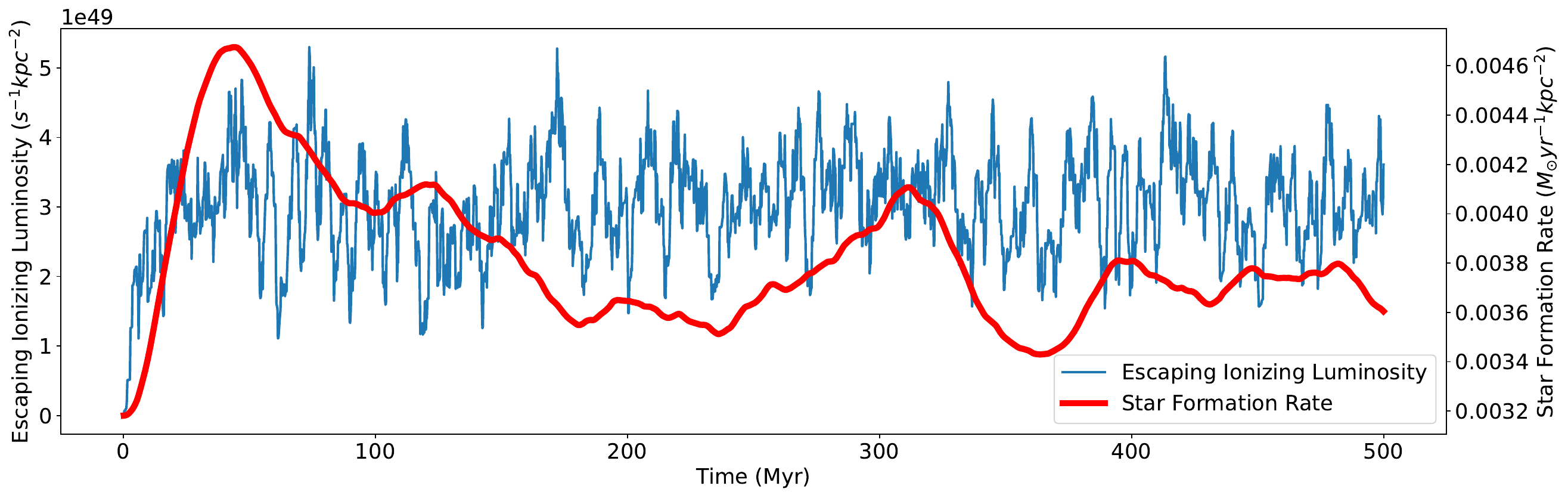}
    \caption{The evolution of total escaping luminosity within the simulation box, and SFR throughout 500 Myr of evolution. SFR is calculated at each step as a function of the amount of mass within 200pc of the midplane. Smoothed luminosity has been averaged over a 10 Myr uniform filter.}
    \label{lumvstime}
\end{figure*}

\begin{figure*}
    \centering
    \includegraphics[width=0.6\textwidth]{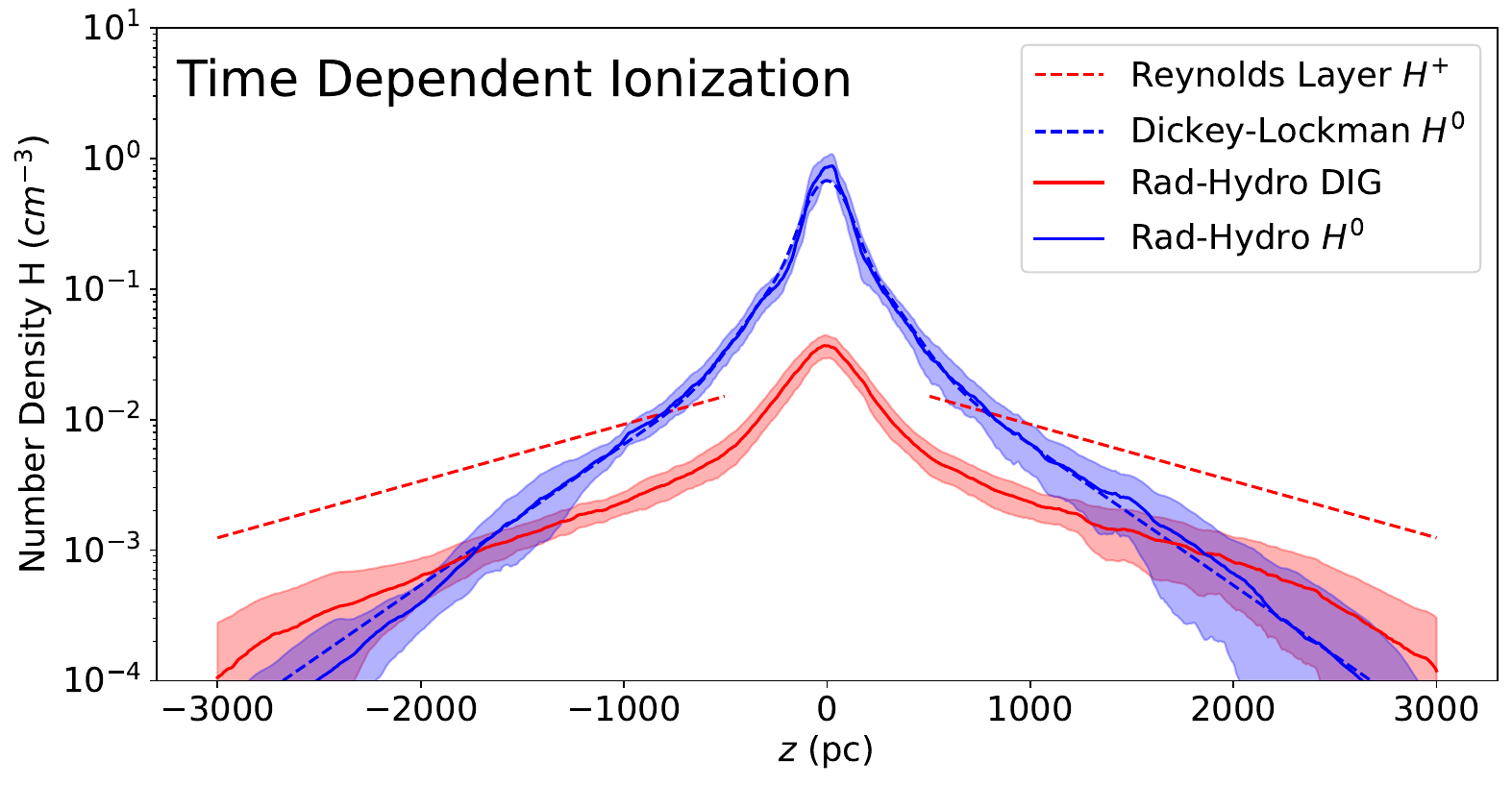}
    \includegraphics[width=0.6\textwidth]{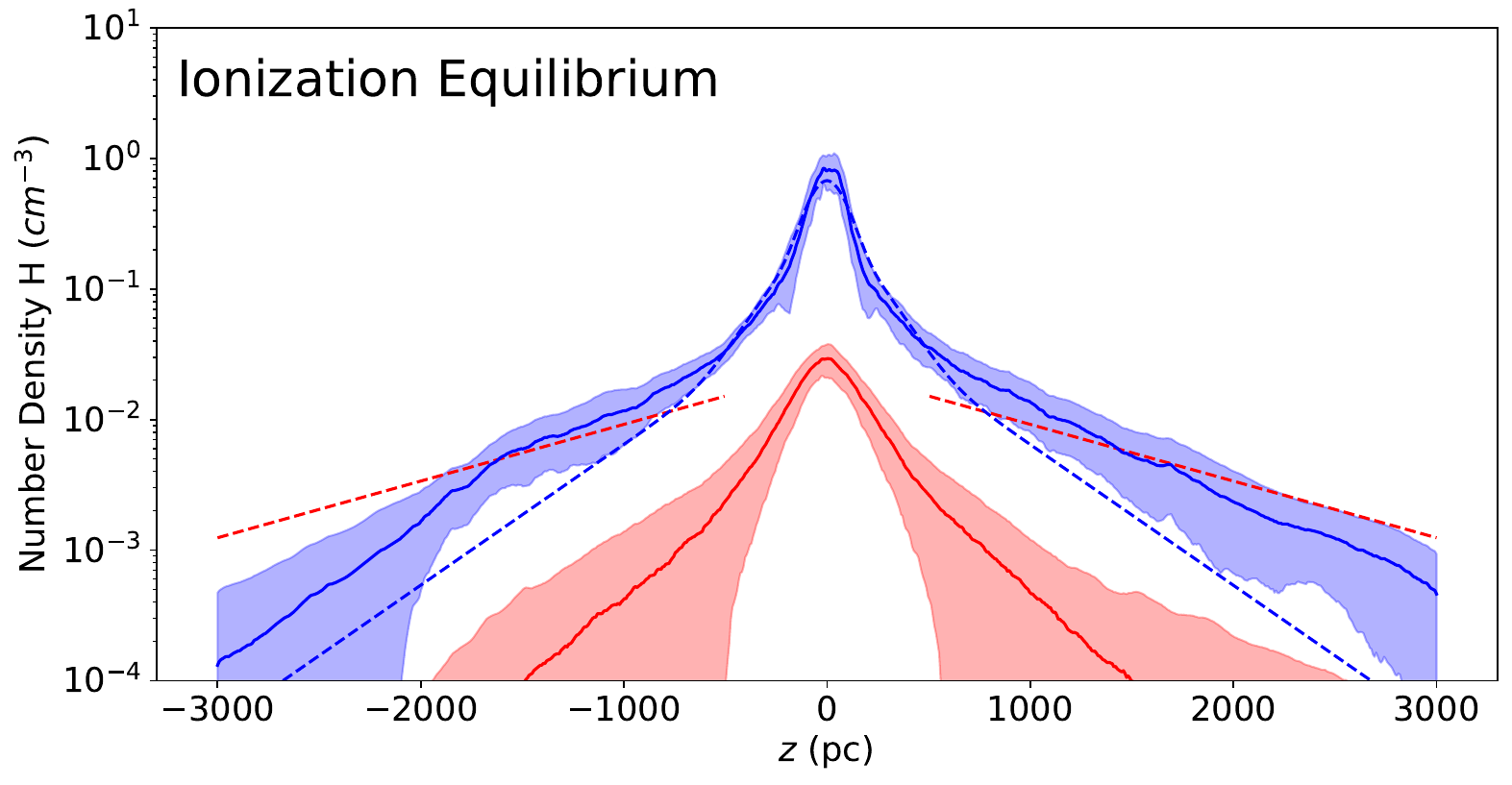}
    \caption{Vertical extent of neutral and warm ionized hydrogen both in the Milky Way and in our fiducial (time-dependent non-equilibrium ionization) and equilibrium-ionization simulations. Solid lines show the median densities throughout the simulation snapshots for times 150~Myr to 500~Myr. Blue dashed lines represent the Dickey-Lockman distribution for neutral hydrogen in the Milky Way, and red dashed lines show the inferred Reynolds layer for ionized hydrogen. The red solid lines in each panel show the warm ionized structure (T < 15000~K) from our simulations, and the solid blue lines show the neutral structure. The upper panel shows results of our fiducial model (with our non-equilibrium ionization scheme), and the lower panel shows the results given the assumption of ionization equilibrium. Shaded regions cover $1\sigma$ intervals. Note that the $1\sigma$ dispersion of the warm ionized gas is much smaller in our time-dependent ionization simulations. Compared to simulations adopting ionization equilibrium, our time-dependent ionization and recombination scheme yields persistent widespread warm ionized gas at higher densities and with less temporal variability. See also figure~\ref{flicker} for images demonstrating the difference in the simulations.}
    \label{squiggle}
\end{figure*}

\begin{figure}
    \centering
    \includegraphics[width=\columnwidth]{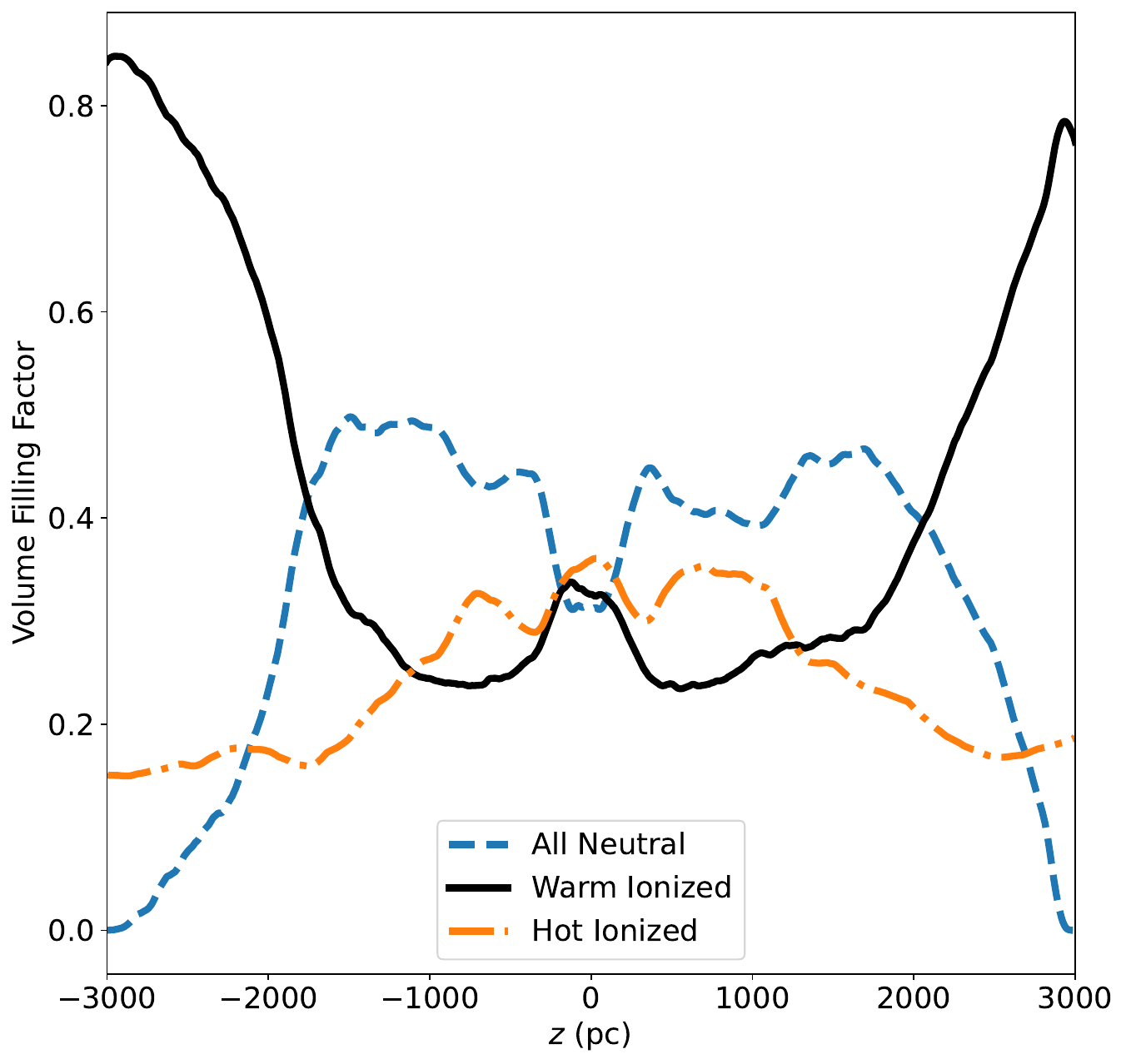}
    \caption{Volume filling factors of the hot ionized medium ($T> 15000$~K), warm ionized medium ($T< 15000$~K), and neutral gas as a function of height in the fiducial model. The lines show mean values for the volume filling factors in our fiducial simulation for all snapshots from 150~Myr to 500~Myr. }
    \label{vff}
\end{figure}

\subsection{Initial conditions}
\subsubsection{Density and temperature structure}

The initial atomic hydrogen number density, $n_H(z)$, (in units of ${\rm {cm}}^{-3}$) was set up as follows: 
\begin{multline}
\label{dl}
    n_H(z) = 0.47e^{-0.5\left(\frac{z}{h_1}\right)^{2}} + 0.13e^{-0.5\left(\frac{z}{h_2}\right)^{2}}  \\
     + 0.077e^{-\left(\frac{|z|}{h_3}\right)} + 0.025e^{-\left(\frac{|z|}{h_4}\right)}
\end{multline}

The first three terms of equation~\ref{dl} represent the `Dickey-Lockman' \citep{dickeylockman} vertical structure of neutral hydrogen in the Milky Way, and the fourth term represents the `Reynolds layer' of ionized hydrogen. The scale heights in kpc are $h_1 = 0.09$, $h_2 = 0.225$, $h_3 = 0.403$, and $h_4 = 1.0$. This density structure was chosen to initialise the simulation box with a structure as close to the expected final quasi-static state for faster convergence. We also hope to avoid unphysically large outflow rates which could occur by initialising with too much gas concentrated towards the midplane. 

We initialise the ionization state of the gas as $99.99\%$ neutral and a temperature of 10,000K. The function described in equation~\ref{dl} is more extended than hydrostatic equilibrium against the external potential, and the simulation will therefore begin by moving into a state of inflow.

\subsubsection{External potential}

An external potential is utilised to simulate the gravity from the stars and gas within the galaxy. The potential used is the same as in \cite{silcc} and \cite{bertkenny}, and the expression for acceleration is:
\begin{equation}
    \boldsymbol{a}= -2\pi G\Sigma_{M}\tanh\left(\frac{z}{b_{M}} \right)
\end{equation}
where $\Sigma$ is the surface density of mass, set as $30 M_{\odot}{\rm pc}^{-2}$, $z$ is height above the midplane and $b_{M}$ is the mass scale height, set as 200~pc.

\section{Results}
\label{results}
\subsection{Summary of models}

Table \ref{maintable} displays the input parameters, and resulting fits for the vertical structure of the DIG in our simulations. All simulations were run on the previously described Cartesian grid with dimensions of $1~\rm{kpc}\times 1~\rm{kpc}\times 6~\rm{kpc}$.

The DIG parameters are retrieved by taking a median warm ionized hydrogen density at each height throughout the time period of 150~Myr to 300~Myr, and fitting to a simple exponential function. Warm gas is selected as having $300K < T < 15000K$. The midplane regions ($|z| < 200~\rm{pc}$) are not included in the fit and neither are regions close to the upper and lower boundaries ($|z| > 2~\rm{kpc}$). The uncertainties are determined from the covariance matrix of each fit, with the standard deviation of density/$\rm{H}\alpha$ at each height being used as the input uncertainties.

\begin{table*}
\resizebox{\textwidth}{!}{%
\begin{tabular}{llllll|lll}
\textbf{Model Name} &
  \textbf{SFR} &
  \textbf{HOLMES Luminosity} &
  \textbf{Escape Fraction} &
  \textbf{Ion State Limiter} &
  \textbf{Resolution} &
  \textbf{DIG $n_{0}$} &
  \textbf{DIG Scale Height} &
  \textbf{$\rm{H}\alpha$ Scale Height} \\
\textbf{} &
  ($M_{\odot} \rm{yr}^{-1}\rm{kpc}^{-2}$) &
  ($s^{-1} \rm{HOLMES}^{-1}$) &
   &
   &
  (pc) &
  ($\rm{cm}^{-3}$) &
  (pc) &
  (pc) \\
Fiducial        & 0.0032  & 0                & 0.1 & Yes & 7.81 & $0.0087\pm 0.0003$ & $766\pm 21$ & $291\pm 6$ \\
Equilbrium      & 0.0032  & 0                & 0.1 & No  & 7.81 & $0.0064\pm 0.0019$ & $343\pm34$ & $197\pm24$ \\
HighSFR         & 0.01    & 0                & 0.1 & Yes & 7.81 & $0.0264\pm 0.0020$ & $442\pm17$ & $227\pm7$ \\
LowSFR          & 0.0005  & 0                & 0.1 & Yes & 7.81 & $0.00020\pm 0.00001$ &$2913\pm 363$ & $782\pm 37$ \\
HOLMESLOW       & 0.0032  & $5\times10^{45}$ & 0.1 & Yes & 7.81 & $0.0203\pm 0.0006$ & $967\pm 25$ & $361\pm 5$ \\
HOLMESMID       & 0.0032  & $1\times10^{46}$ & 0.1 & Yes & 7.81 & $0.0440\pm 0.0014$ & $633\pm 11$ & $293\pm 4$ \\
DIM             & 0.0032  & 0                & 0.02& Yes & 7.81 & $0.0063\pm 0.0002$ & $1183\pm 29$ & $422\pm 7$ \\
BRIGHT          & 0.0032  & 0                & 0.5 & Yes & 7.81 & $0.0330\pm 0.0025$ & $517\pm 16$ & $226\pm 6$ \\
NOPHOTONS       & 0.0032  & 0                & 0.0 & Yes & 7.81 & $0.0041\pm 0.0001$ & $1850\pm 81$ & $569\pm 15$ \\
\end{tabular}%
}
\caption{Table of main simulation parameters and resulting vertical structures of ionized hydrogen. DIG is defined as ionized gas with $300K < T < 15000K$. }
\label{maintable}
\end{table*}

\subsection{Convergence}
\subsubsection{Radiation timestep}
\label{radconv}

In order to ensure the frequency of the radiation step (and hence frequency of ionization state update) is sufficient to converge the results for our purposes, two simulations were run, identical in every parameter except the radiation timestep. The coarse timestep model is run at the fiducial radiation timestep of every 0.2 Myr of simulation time, with the fine timestep test running the radiation step every 0.05 Myr. Figure \ref{radconvfig} shows the results of these two simulations averaged between 150 and 500 Myr of evolution by taking a median value for ionized and neutral density for each height.

It can be seen that the two simulations are generally in agreement, with some differences due to the inherent stochasticity in our star formation algorithm. Throughout the evolution there are some differences at any given time as each simulation goes through periods of inflow and outflow which are not in sync with each other. However, it is not expected that these differences would affect our conclusions regarding the importance of including time-dependent ionization to produce and maintain widespread vertically extended diffuse ionized gas.

\subsubsection{Spatial resolution}

The closely related work of \cite{kimostriker17} carried out resolution convergence tests throughout a similar suite of simulations, and found that a resolution of $\Delta x \leq 16 \rm{pc}$ was sufficient to converge results. Our simulations with $\Delta x = 7.81 \rm{pc}$, attain this resolution. Further exploration into resolution effects represents future work expanding the capabilities of the hydrodynamics code.

\subsection{Fiducial model}
\label{fiducial}
Our fiducial model has time-dependent ionization calculations turned on, an initial SFR surface density of $0.0032 M_{\odot} \rm{yr}^{-1} \rm{kpc}^{-2}$, and no ionizing radiation from hot evolved stars.

\subsubsection{Density and ionization structure}

The initial conditions represent a structure which is more extended than can be supported in hydrostatic equilibrium. As we start with no sources in the simulation box, this means that the first $\sim50$~Myr of evolution is dominated by inflow as the first sources start to form. The extended structure is broadly neutral at the start of this period, with the ionized gas being confined to mostly spherical H{\sc ii} regions within the undisturbed, smooth ISM. The occasional type Ia supernova occurs at higher altitudes than the H{\sc ii} regions (up to around 500~pc), opening up large ($\sim200$~pc radius) bubbles of hot ($\sim1\times10^{8}$~K) gas which are subsequently swept up in the inflowing material and fall towards the midplane. 

As more material flows in, the vertical density distribution sharpens into a higher density disc which leads to an increased SFR. Beyond $\sim30$~Myr, the discrete, spherical HII regions become less easily distinguishable from the widespread diffuse ionized component as the uniform structure is disturbed by the first type II supernovae close to the midplane. As the ISM is still in a state of inflow at this time, the high altitude gas remains neutral, as it is unaffected by the midplane star formation. For the gas to be ionized, it must be either irradiated by ionizing photons (photoionized) or heated by supernovae (collisionally ionized). The photon transport is inhibited in uniform media. Turbulent outflows or turbulence due to supernovae are crucial in the support of ionized gas at altitude. This means that the high altitude gas remains neutral until the simulations develop outflows. 

At approximately 75~Myr, the balance between gravitational inflow and outward support due to supernovae and ionizing photons changes, and a turbulent mix of warm/hot gas is driven to high altitudes. As the high altitude gas is heated and loses its initial uniformity, the vertical extent of the ionized gas increases rapidly, with the hot gas having been mostly thermally collisionally ionized and the warm gas primarily ionized by ionizing photons whose transport has been permitted by the low density channels which have opened up in the turbulent ISM.  Figure \ref{bigplot} shows the structure of a single snapshot out to $\pm 2$~kpc height at a time of 350~Myr, and Figure \ref{panelevo} shows the first 150~Myr of evolution of edge on projected column density.

Looking at the evolution of the star formation rate throughout the simulation in figure \ref{lumvstime}, we see the SFR increases steadily as the simulation begins in a state of inflow, increasing the midplane mass and hence the SFR. The SFR continues to vary throughout the simulation as the evolution goes through periods of net outflow and inflow from and into the disc. These periods of inflow and outflow are also noted in the simulations of \cite{silcc} and \cite{kadofong}.

Beyond 300~Myr of evolution, the periods of inflow and outflow of gas become out of sync between the north and south sides of the midplane. This leads to many snapshots with large asymmetries comparing the structure above and below the midplane. The periods of inflow and outflow continue for the entire 500~Myr of the simulation. In the later stages of the simulation we observe instabilities in the dense disc, whereby the disc `wobbles' above and below $z = 0$. With the disc oscillating above and below the midplane of the simulation grid, this introduces further asymmetries in ionizing flux escaping the midplane into either side of the disc.

Figure \ref{squiggle} shows the median density structure of neutral and ionized hydrogen at each value of $z$ for our fiducial simulation. The displayed line is that of the median density for every value of $z$. Our results are plotted alongside the `Dickey-Lockman' estimation for the vertical distribution of neutral hydrogen in the Milky Way, and separately plotted is the `Reynolds layer' estimation for the vertical distribution of ionized hydrogen. The median values are shown at each height for every snapshot in the range 150~Myr to 500~Myr, with $\pm 1\sigma$ fill.

It can be seen in the upper panel of figure \ref{squiggle} that this fiducial model with OB star photoionization and supernova feedback and a time-dependent ionization calculation produces vertical structures of neutral hydrogen similar to those inferred for the Milky Way. The ionized gas layer produced is of a similar scaleheight to observed, but at slightly lower densities. By defining the DIG as gas which is ionized with temperatures in the range 5000~K to 15000~K, this model produces a DIG layer which fits an exponential disc of scale height 766~pc between heights of 200~pc and 2~kpc.

Figure \ref{vff} shows the volume filling factors of the hot ionized medium ($T>15000$~K), warm ionized medium ($T<15000$~K) and total neutral medium (warm neutral plus cold neutral) in the fiducial model. We do not distinguish between the warm and cold neutral medium because our simulations do not include photoelectric heating from dust grains, an important heating mechanism for the warm neutral medium. We find that the midplane volume filling factor of the DIG is $\sim30$\%, increasing to $\sim80$\% at the top of our simulation box.

For $|z|< 250$~pc, we find the volume filling factors and mass fractions as displayed in table \ref{vfftable}, again without distinguishing between cold and warm neutral medium. Also displayed are observational estimates from \cite{tielens05}, and the values resulting from the full SILCC simulation of \cite{rathjen21}. 

The midplane volume filling factor of WIM/DIG in our fiducial model is close to the observational estimate from \citet{tielens05} and the SILCC simulations from \citet{rathjen21}. Similar to the SILCC simulations our model underestimates the midplane mass fraction of the DIG compared to observational estimates.

\begin{figure*}
    \centering
    \includegraphics[width=\textwidth]{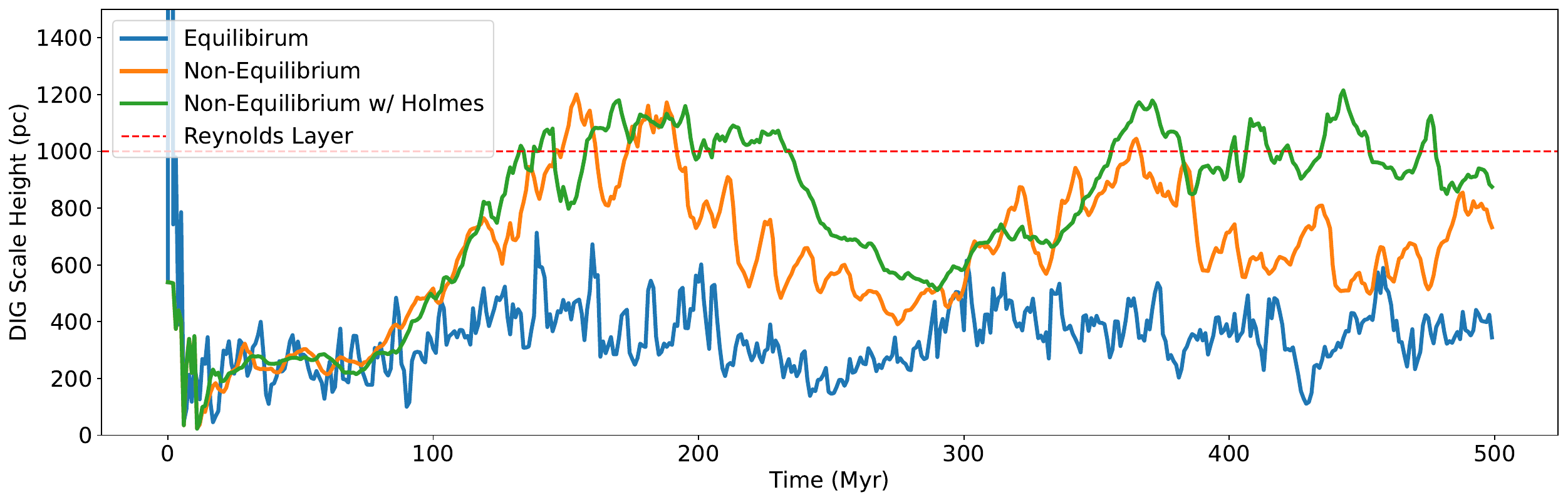}
    \caption{Time evolution of scaleheight of DIG layer in simulations fiducial, equilibrium and HOLMESLOW. Scaleheights are derived from fitting to decreasing exponential. Heights of $|z|<500\rm{pc}$ and $|z|> 2000\rm{pc}$ are not included in the fit. DIG is defined as ionized gas with $T < 15000\rm{K}$.}
    \label{scaleheightevo}
\end{figure*}

\begin{table*}
    \centering
    \begin{tabular}{c|c|c|c|c|c|c|c}
        Component & $\bar{n}_{CMacIonize}$ (${\rm cm}^{-3}$) & $VFF_{CMacIonize}$ & $VFF_{SILCC}$ & $VFF_{Teilens}$ & $MF_{CMacIonize}$& $MF_{SILCC}$& $MF_{Tielens}$\\
        \hline \\
        HIM     & 0.0016 & 0.38 & 0.35 & 0.5 & 0.004 & 0.0003 & -  \\ 
        WIM     & 0.024 & 0.35 & 0.14 & 0.25 & 0.056 & 0.06 & 0.14\\
        Neutral & -     & 0.27 & 0.52 & 0.31 & 0.94 & 0.94 & 0.86\\
    \end{tabular}
    \caption{Mean volume filling factors (VFF), mean densities ($\bar{n}_{CMacIonize}$) and mass fractions of each component of the ISM in our fiducial model, as averaged throughout the region $|z|<250$~pc. Values are medians from all snapshots between 150~Myr and 500~Myr. Also shown are results of the flagship SILCC simulation from \citet{rathjen21}, and observational estimates from \citet{tielens05}. Note that the definition of 'warm' gas in \citet{rathjen21} is $300K < T < 3\times 10^{5} K$, whereas in this work is $300K < T < 1.5\times 10^{4}K$. }
    \label{vfftable}
\end{table*}

\subsubsection{Pulsar Dispersion versus H$\alpha$ Scale Height}

Two commonly used observational techniques to measure DIG scale height are pulsar dispersion measures, and the spatial distribution of H$\alpha$ emission. The former is proportional to the total integrated electron density whereas the latter is proportional to integrated electron density squared. H$\alpha$ scale heights are hence often simply doubled to infer a DIG/free electron scale height following an assumption of a constant volume filling factor \citep{hill14,dk17}. It can be seen in table~\ref{maintable} however, that in all simulations the H$\alpha$ scale height is less than half the DIG scale height. This can be explained by a rising DIG volume filling factor with height, which is evident in the fiducial model in figure~\ref{vff}. Note we are using DIG scale height and free electron scale height interchangeably here, due to the negligible mass fraction of the HIM as seen in both observations and simulations in table~\ref{vfftable}. From the pulsar dispersion measure results of \citet{ocker2020}, the local free electron scale height has been measured as $1.57^{+0.15}_{-0.14}$~kpc, with a midplane density of $0.015\,{\rm cm}^{-3}$. Using our simulation HOLMESLOW as the closest DIG layer in vertical structure to these observed values, we find the H$\alpha$ derived electron scale height of 722~pc to be approximately 75\% of the true DIG scale height in that simulation. 


\subsection{Ionization Equilibrium}

To study the importance of non-equilibrium ionization of the DIG, the fiducial run is repeated without the ionization state limiters as described in section \ref{ionstatelimit}. This forces the gas to be in ionization equilibrium at all locations and times in the simulation. Whenever ionized gas becomes either shadowed by dense clumps, or a source turns off, the gas instantaneously becomes neutral.

The dynamical evolution in this simulation is similar to that of the fiducial model. Looking at the distribution of total gas mass, the simulation is similar in general structure. However the ionized fractions above heights of 1~kpc are vastly different in the equilibrium run, and the ionized hydrogen mass at height is very time variable.

Due to the steep slope of stellar mass versus ionizing luminosity, and the relative rarity of stars of mass $>80M_{\odot}$, the total ionizing luminosity within the box is at any time dominated by a very small number of stars. These stars also have the shortest lifetimes of < 3.5 Myr, so appear and disappear within a few snapshots in the simulations. This means the total ionizing luminosity in the box can change very rapidly (see figure \ref{lumvstime}).

There are also projection effects whereby fast moving material driven by supernovae near the midplane can quickly move across the line of sight between any single bright ionizing star and a very large volume of gas. The stochasticity in the escaping ionizing luminosity combined with these projection effects without a limiter on rate of change of ionization state gives a flickering effect to the ionized gas structure.

The assumption of ionization equilibrium also leads to a decrease in the total amount of ionized gas in the simulation. Figure \ref{squiggle} displays the neutral and ionized hydrogen densities versus height above and below the midplane adopting ionization equilibrium. Again we are showing the median value at each height for every snapshot in the range 150~Myr to 500~Myr, with $\pm 1\sigma$ fill.

Figure~\ref{scaleheightevo} shows the time evolution of the DIG scaleheight in the three following runs, fiducial, equilibrium and HOLMESLOW. Again it can be seen that the equilibrium run produces much smaller and more variable scaleheights than the run utilising the full non-equilibrium calculation. The DIG scaleheight is seen to increase and decrease in the fiducial run as the simulation enters periods of outflow and inflow. The period of this oscillation is estimated to be on the scale of approximately 200 Myr.

\begin{figure*}
    \centering
    \includegraphics[width=\textwidth]{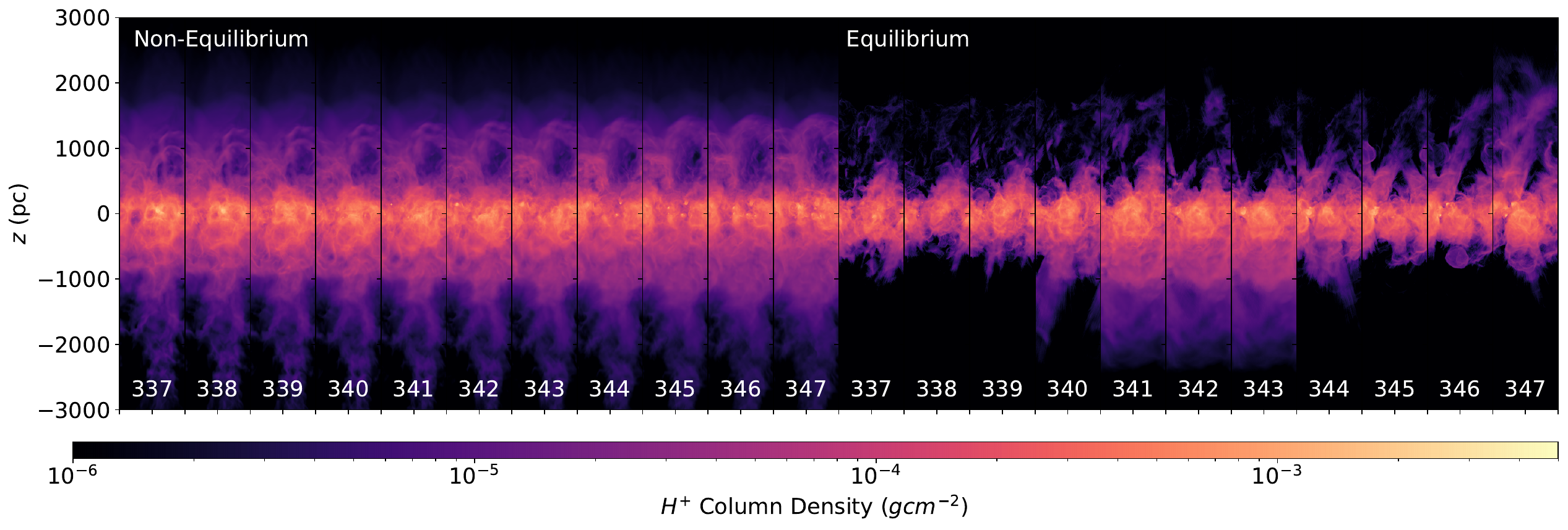}
    \caption{An illustration of the `flickering' effect in the ionization equilibrium simulation. Panels show the edge-on column density of ionized hydrogen at various times between 337~Myr and 347~Myr. The leftmost 11 panels (labelled `Non-Equilibrium') show the results of our fiducial (time dependent non-equilibrium ionization) simulation. The rightmost 11 panels (labelled `Equilibrium') show the ionization equilibrium simulation. White text at the bottom of each panel shows the simulation time in Myr. Note the extent and stability of the ionized gas in the non-equilibrium (time dependent) ionization simulation. The flickering effect in the ionization equilibrium simulation is evident between 341 and 343 Myr where the ionized gas is briefly much more extended.}
    \label{flicker}
\end{figure*}

Figure~\ref{flicker} shows a time series of snapshots of the column density of ionized gas for the fiducial simulations with and without non-equilibrium ionization. The non-equilibrium ionization simulation gives a more vertically extended distribution of ionized gas that is largely unchanged during the 10~Myr of evolution shown. In contrast, the simulation where ionization equilibrium is adopted leads to less ionized gas at high altitudes and more variability. Note in particular the transient increase in the volume of ionized gas between times 341~Myr and 343~Myr.

In summary, we agree with the findings of \citet{kadofong} that when photoionization equilibrium is assumed the resulting DIG is highly variable and not as extended as in the Milky Way. The inclusion of non-equilibrium ionization in our simulations maintains the DIG at high altitudes resulting in neutral and ionized density profiles similar to those inferred for the Milky Way.

\subsection{Hot evolved low mass stars}

\begin{figure}
    \centering
    \includegraphics[width=\columnwidth]{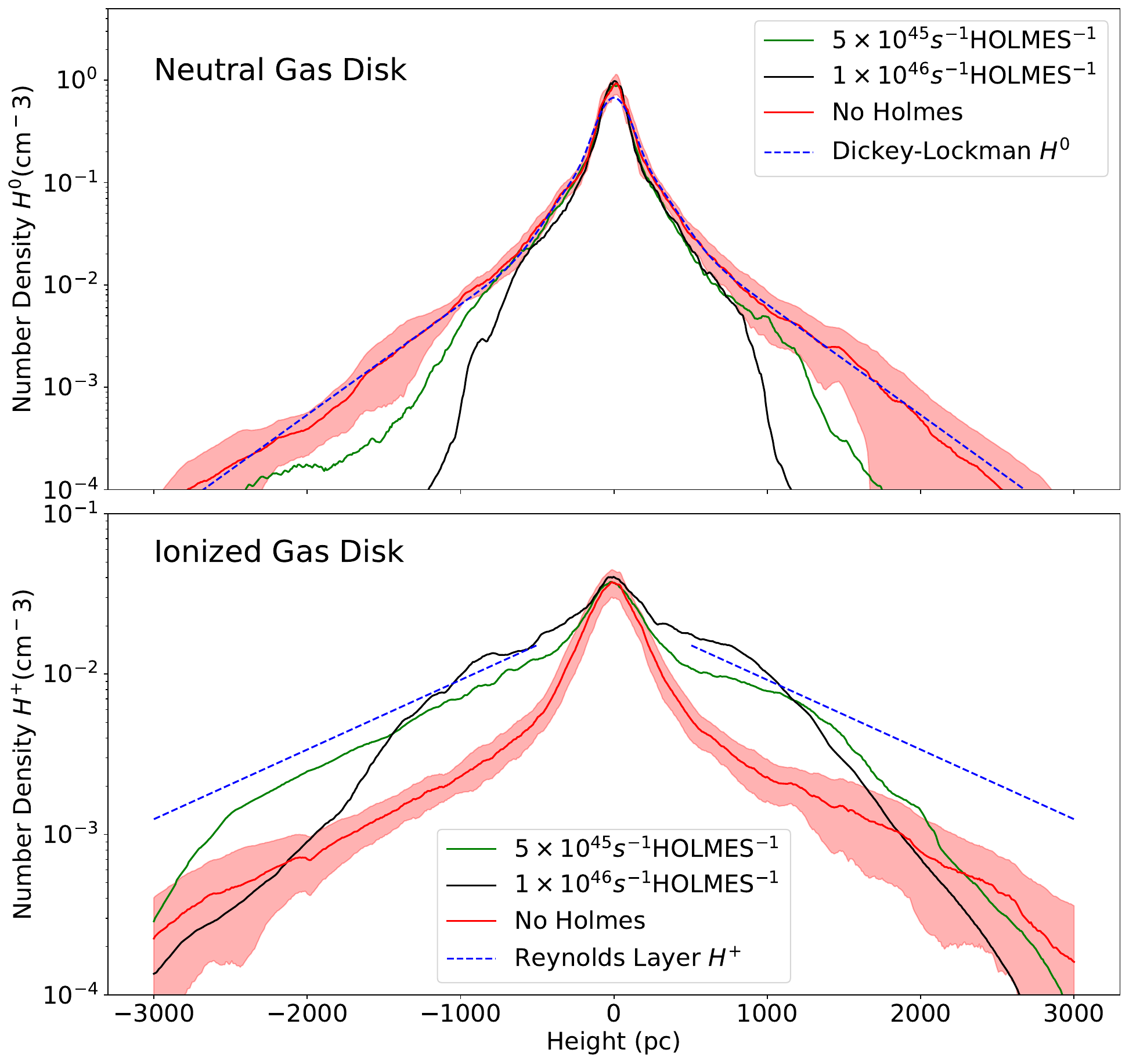}
    \caption{Vertical extent of neutral and ionized hydrogen both in the Milky Way and in our HOLMES simulations. Solid lines show the median densities throughout the simulation snapshots. Lower panel shows ionized hydrogen and upper panel is neutral. Dashed blue lines show the inferred neutral and ionized density structure for the Milky-Way, solid lines represent simulations with different HOLMES luminosities. Lines are median value for each height throughout all snapshots from 150-300~Myr. Red fill on each panel shows $1\sigma$ variation around the control run with no HOLMES.}
    \label{holmessquiggles}
\end{figure}

Figure \ref{holmessquiggles} displays the results of the three simulations including hot evolved low mass sources of differing ionizing luminosities, alongside the fiducial run of the same SFR. From figure \ref{lumvstime} we see that the mean total escaping ionizing luminosity is approximately $3.1\times10^{49}\,{\rm s}^{-1}$. With this representing 10\% of the total ionizing luminosity due to the 0.1 escape fraction, the ionizing luminosity of the low mass stars in HOLMESLOW and HOLMESMID comprises 2.0\% and 3.8\% of the available ionizing photon budget.

Compared to the Dickey-Lockman distribution, the fiducial run has more neutral gas at large altitudes, and in these regions produces less ionized hydrogen than the Milky Way's Reynolds layer. Figure~\ref{holmessquiggles} demonstrates that more DIG is produced with increasing ionizing luminosity from the low mass stars. The simulation with a total ionizing luminosity from low mass stars of $5\times10^{48} {\rm s}^{-1}\rm{kpc}^{-2}$ (approximately 2.0\% of total available) most closely matches the Reynolds layer.

\begin{figure}
    \centering
    \includegraphics[width=\columnwidth]{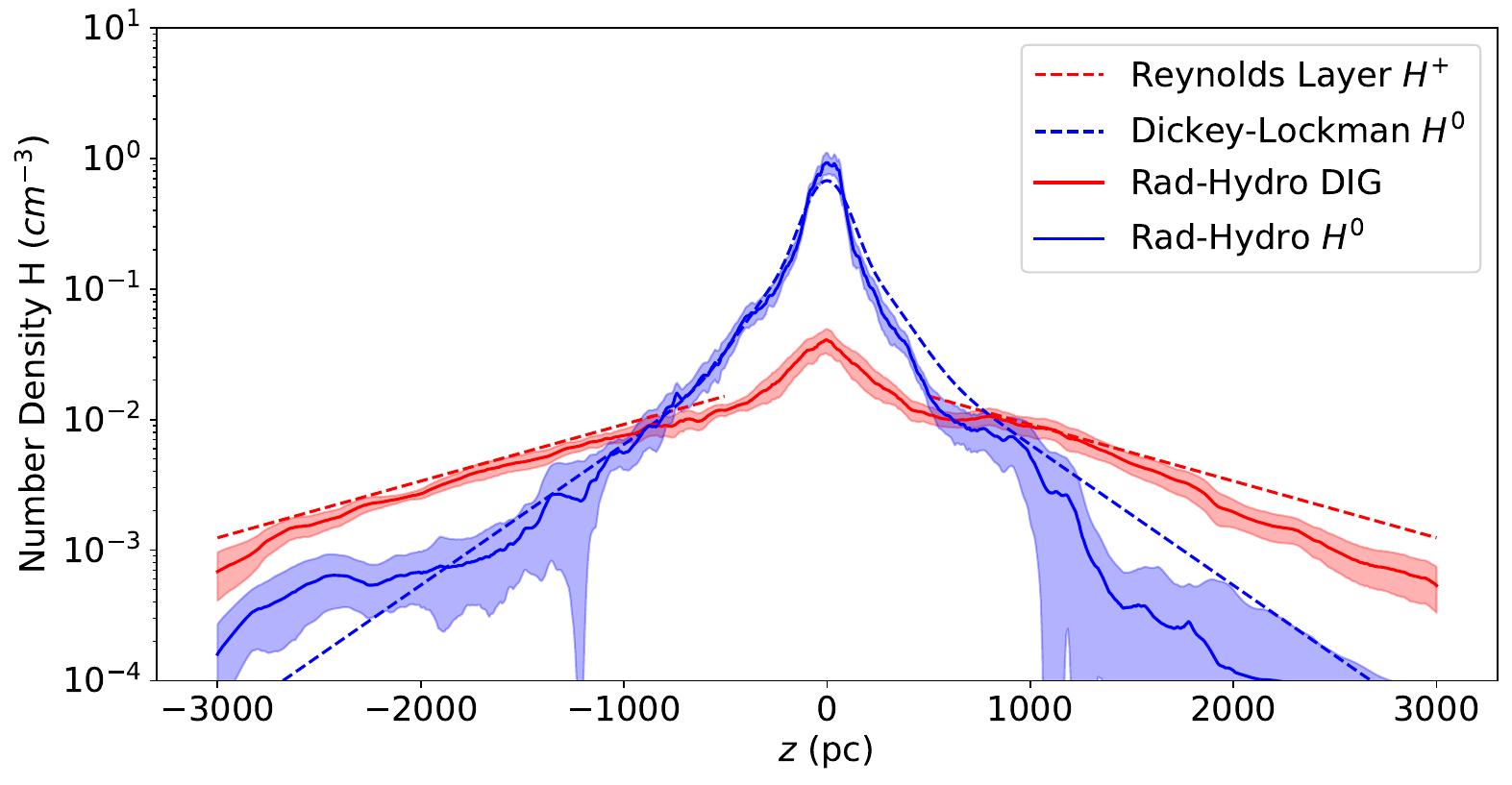}
    \caption{Vertical extent of neutral and warm ionized hydrogen both in the Milky Way and in our HOLMESLOW simulation averaged from 150-220~Myr. Lines are the same as figure~\ref{squiggle}.}
    \label{milkywaybest}
\end{figure}

Also displayed in figure~\ref{scaleheightevo} is the time-evolution of the DIG scaleheight in the HOLMESLOW simulation. While both the fiducial and HOLMESLOW run oscillate between states of high and low DIG scaleheight, the simulation including a dim population of HOLMES sources is seen to support a DIG layer with a higher scaleheight for longer, and one which recovers faster after a period of inflow. The HOLMESLOW simulation appears to match best the Milky Way vertical structures, and for periods of around 70~Myr, the median structure is very close to expected densities. Figure~\ref{milkywaybest} shows the median structure from 150-220~Myr in the HOLMESLOW simulation which is very close Milky Way structures.

\subsection{Effect of star formation rate}

Our fiducial model was initialised such that the SFR within the box was close to the mean Milky Way value ($3.2\times 10^{-3}M_{\odot}\, {\rm{yr}}^{-1}\, {\rm{kpc}}^{-2}$), and then allowed to change in proportion to the gas mass below a height of 200~pc as $M_{\rm gas}^{1.4}$. This initial SFR sets the normalisation of the Kennicutt-Schmidt relation (equation \ref{kennicutt}). We note from the work of \cite{mwsfr} that the SFR surface density varies throughout the Milky Way by one to two orders of magnitude. We therefore ran two further simulations representing a low and high star formation rate of $5\times 10^{-4}M_{\odot}\, {\rm{yr}}^{-1}\, {\rm{kpc}}^{-2}$ and $1\times 10^{-2}M_{\odot}\, {\rm{yr}}^{-1}\, {\rm{kpc}}^{-2}$.

\begin{figure*}
    \centering
    \includegraphics[width=\textwidth]{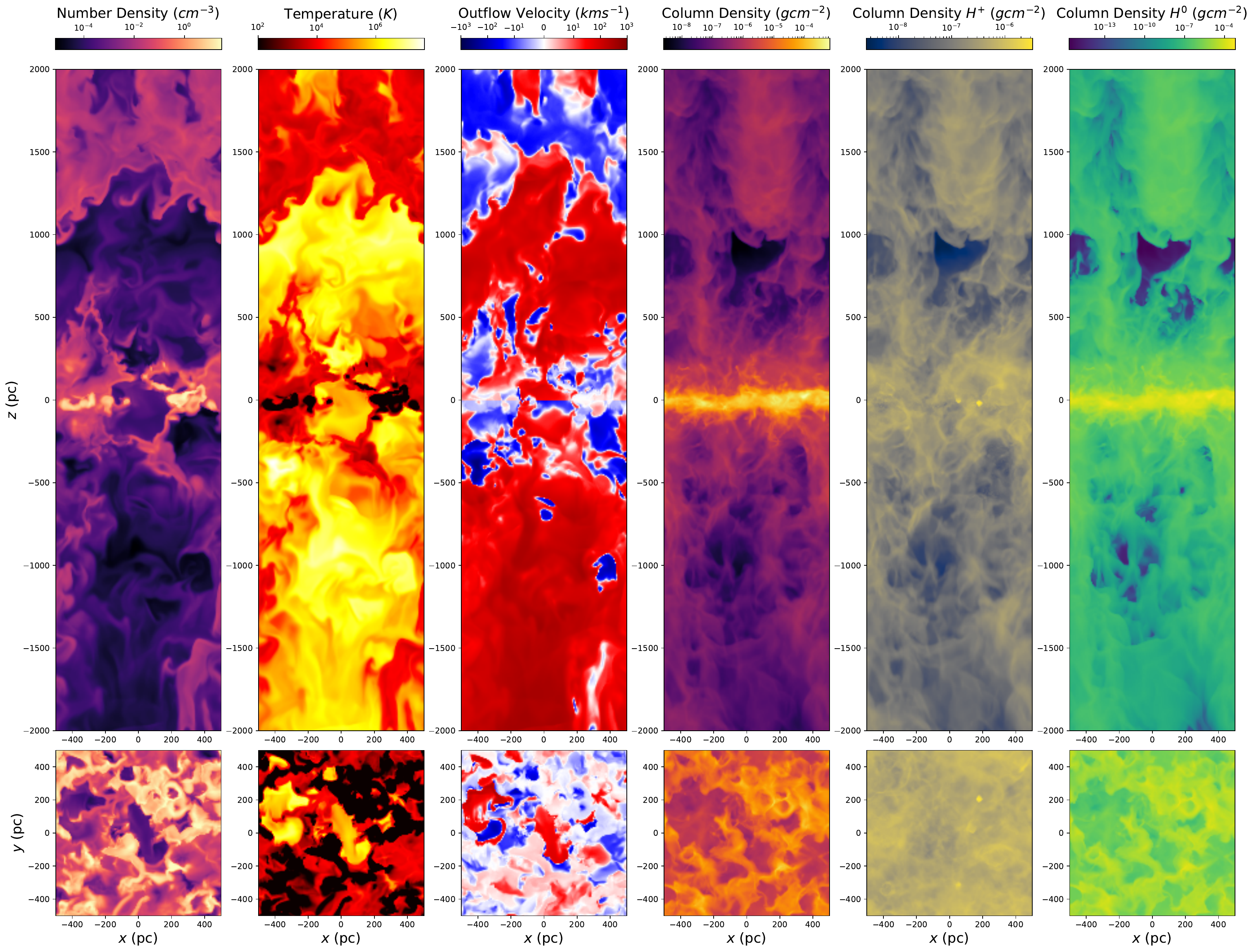}
    \caption{A visualisation of a snapshot from the low SFR model at a time of 350~Myr. Columns are the same as figure \ref{bigplot}.}
    \label{lowsfrplot}
\end{figure*}

\begin{figure*}
    \centering
    \includegraphics[width=0.6\textwidth]{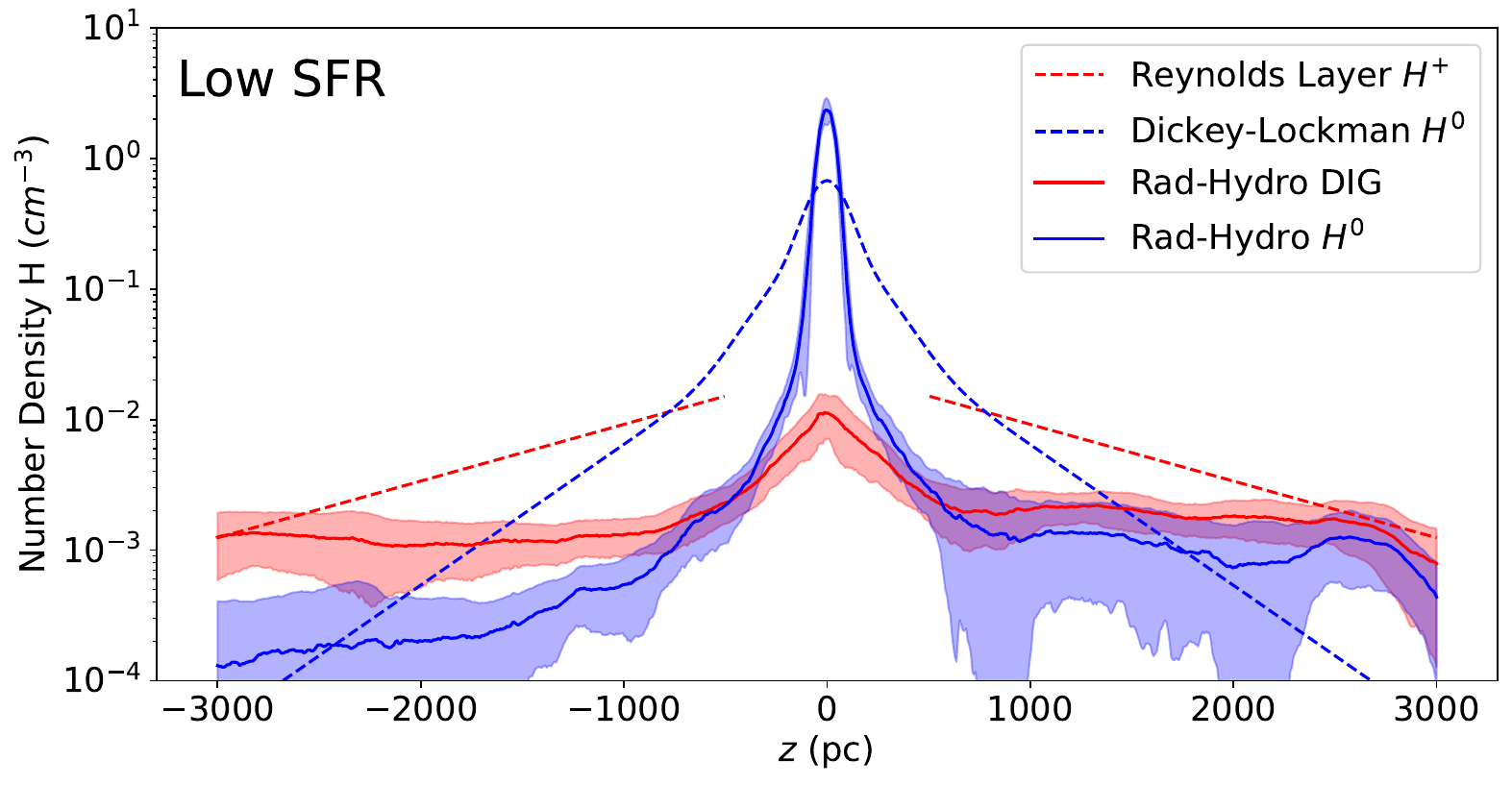}
    \caption{Vertical extent of neutral and ionized hydrogen both in the Milky Way and in our low SFR simulation. Lines and shaded regions are the same as in figure \ref{squiggle}. Results are median value for all snapshots from 150~Myr to 400~Myr. Note that the neutral material is much more sharply confined to the midplane, and there is less material supported at height.}
    \label{lowsfrsquiggle}
\end{figure*}

We find that in the low SFR model, the initial state of inflow continues for longer, causing a sharper peak in density around the midplane which persists throughout the simulation. The extended structure is at much lower density than the fiducial model. Outflows are inhibited due to the low supernovae rate, and high altitude gas is almost entirely ionized due to the low densities. Figures \ref{lowsfrplot} and \ref{lowsfrsquiggle} show the structures formed with this low SFR simulation.

\begin{figure*}
    \centering
    \includegraphics[width=\textwidth]{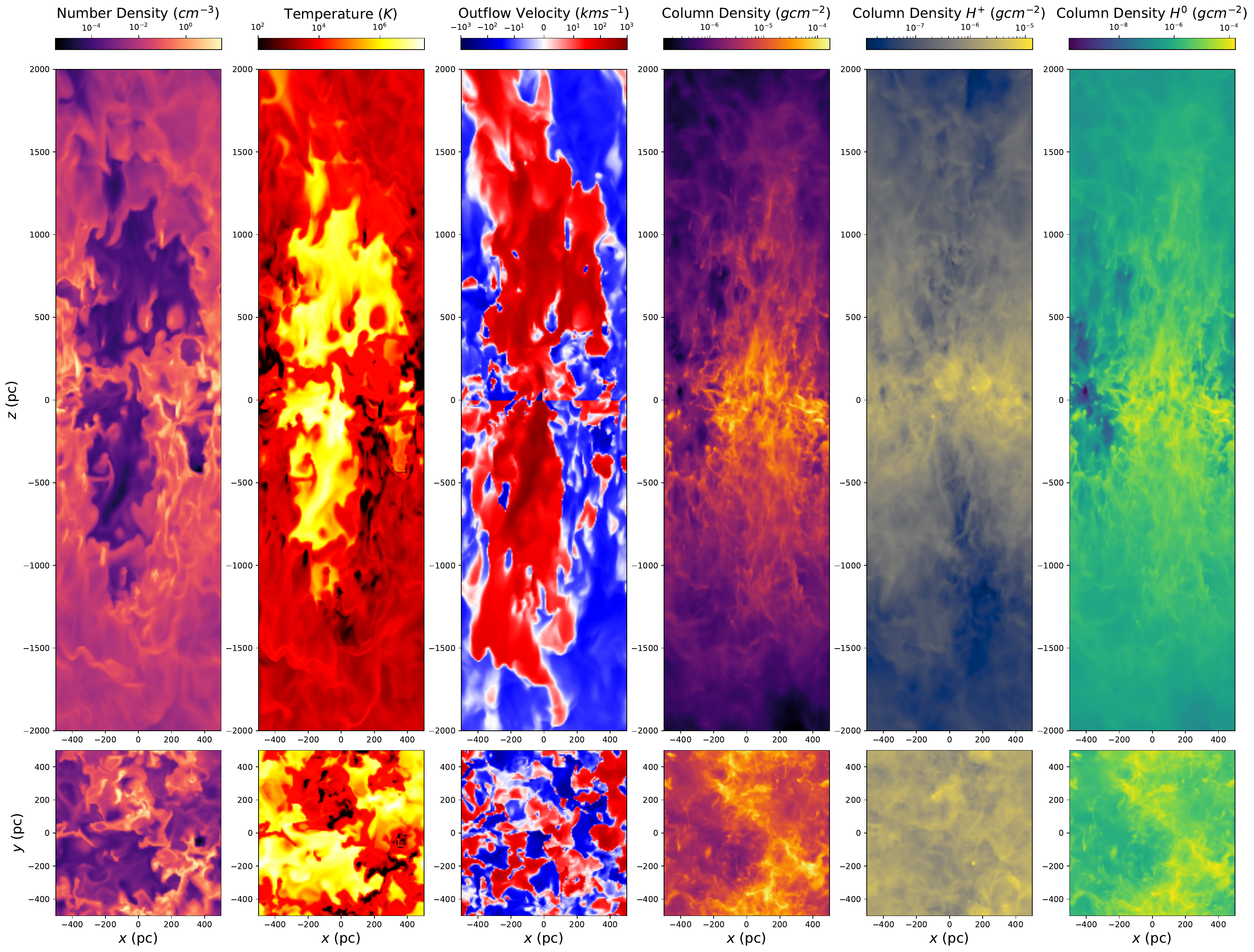}
    \caption{A visualisation of a snapshot from the high SFR model at a time of 350~Myr. Columns are the same as figure \ref{bigplot}. Note that the disc is almost completely disrupted and there is much more material at high altitudes.}
    \label{highsfrplot}
\end{figure*}

\begin{figure*}
    \centering
    \includegraphics[width=0.6\textwidth]{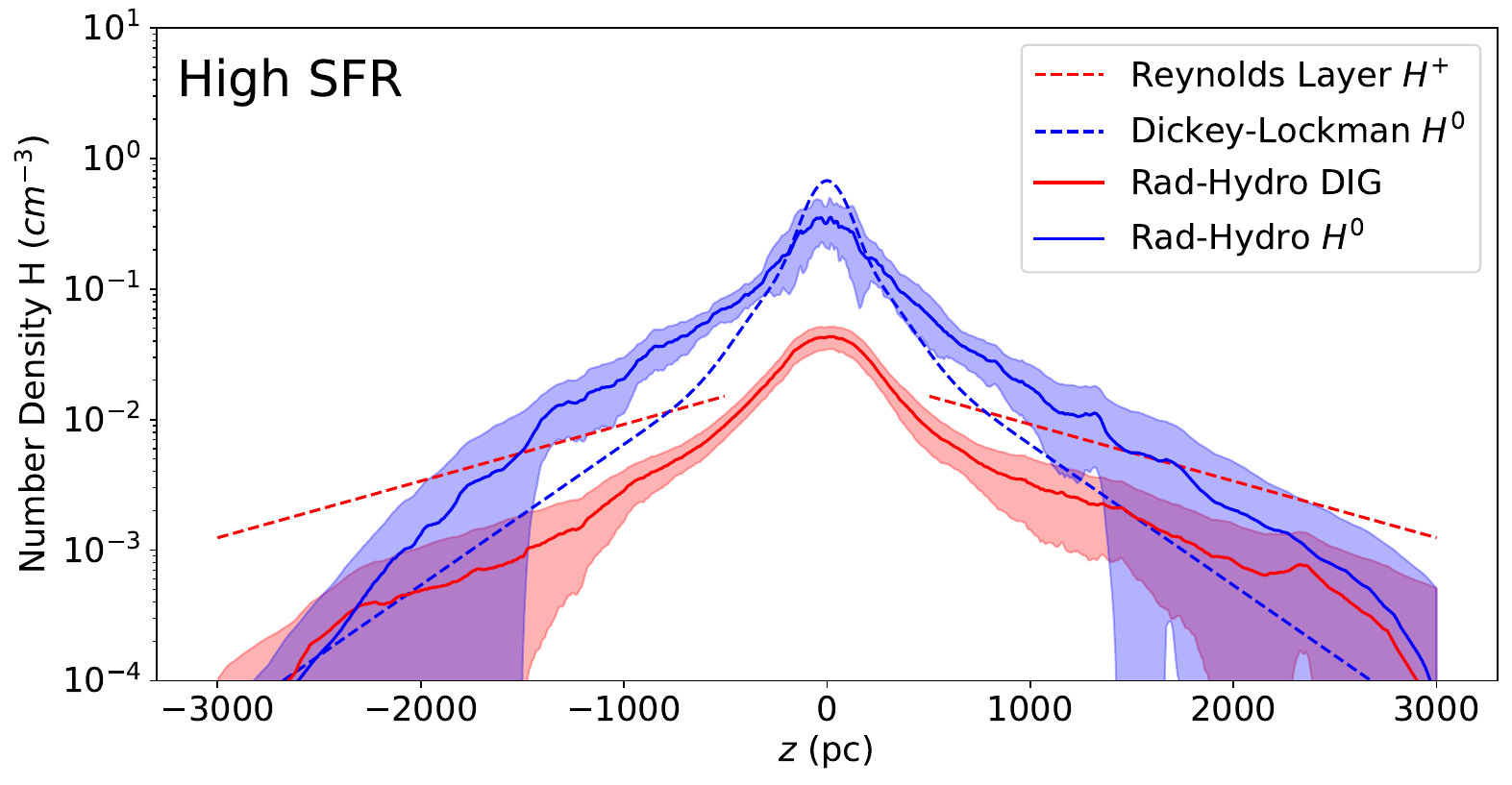}
    \caption{Vertical extent of neutral and ionized hydrogen both in the Milky Way and in our high SFR simulation. Lines and shaded regions are the same as in figure \ref{squiggle}. Results are median value for all snapshots from 150~Myr to 400~Myr. Note that the neutral disc has been destroyed near the midplane, and there is an increased amount of material at heights of 500~pc to 2~kpc.}
    \label{highsfrsquiggle}
\end{figure*}

In the high SFR simulation, the ISM does not collapse into a sharp disc in the same manner as with the two lower SFR runs. For the first 200~Myr we see a mostly stable structure at height, with gas densities comparable to those inferred for the Milky Way. The DIG density is still lower than observed, despite the higher luminosity sources towards the midplane. As the simulation evolves further, the disc is disturbed in the midplane and the disc density decreases below expected levels with the high supernova rate contributing to more powerful outflows. Figures \ref{highsfrplot} and \ref{highsfrsquiggle} show the structures formed with the high SFR simulation.

The shape of the neutral disc structure is clearly sensitive to the SFR. Low SFR simulations produce discs that are higher density, more sharply peaked towards the midplane and less vertically extended. High SFR simulations produce discs that are more vertically extended with lower peak densities, and potentially completely disrupted discs.

\subsection{Effect of Ionizing Photon Escape Fraction}
\begin{figure}
    \includegraphics[width=\columnwidth]{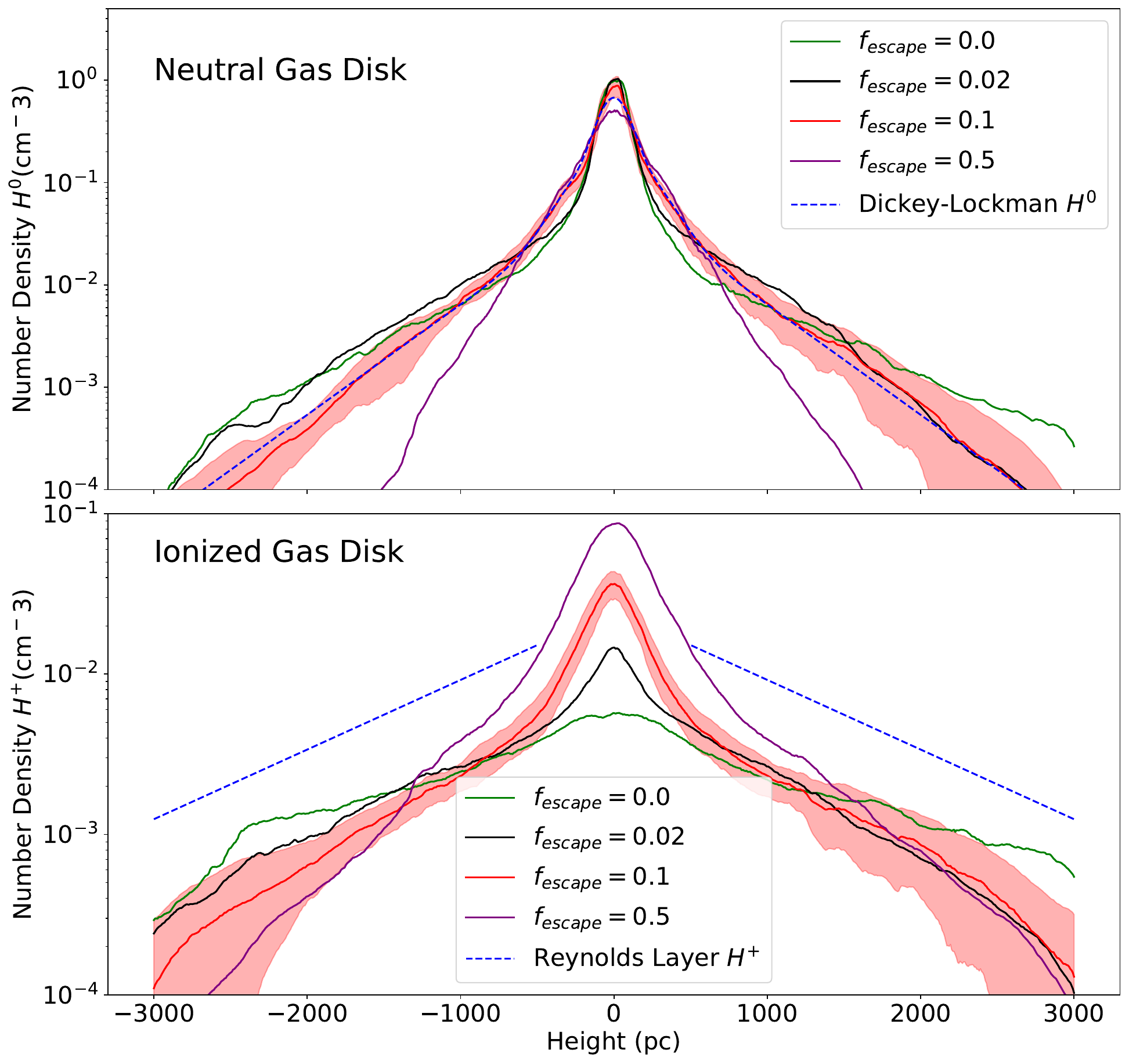}
    \caption{Vertical extent of neutral and ionized hydrogen both in the Milky Way and in our different escape fraction simulations. Solid lines show the median densities throughout the simulation snapshots. Lower panel shows ionized hydrogen and upper panel is neutral. Dashed blue lines show the inferred neutral and ionized density structure for the Milky-Way, solid lines represent simulations with different escape fractions. Lines are median value for each height throughout all snapshots from 150-500 Myr. Red fill on each panel shows $1\sigma$ variation around the fiducial run with escape fraction of 0.1.}
    \label{leakageplot}
\end{figure}

With the ionizing photon escape fraction from unresolved molecular clouds being treated as an unknown, we have run 4 simulations to investigate this parameter space. In table~\ref{maintable} these runs are BRIGHT, fiducial, DIM, and NOPHOTONS, with escape fractions of 0.5, 0.1, 0.02, and 0.0 respectively. 
Figure~\ref{leakageplot} shows the vertical distribution of neutral and ionized hydrogen in these 4 simulations. Whilst these simulations are initialised as identical runs with only the escape fraction varying, the star formation algorithm varies SFR with surface density, and hence the runs with more efficient clearing out of the midplane (brighter runs) tend to have a lower average SFR. This lower SFR in brighter runs is seen to decrease the strength of outflows. This is similar to the that of \citet{rathjen21}, who found that including feedback mechanisms such as photoionization would shut down star formation in their sink particle algorithm, and inhibit the outflow of gas to high altitudes. This highlights the importance of the dynamical effect of ionizing photons, as is shown by the simulations of \citet{bertkenny}, which exclude the effect of supernova feedback.

\section{Discussion}
\label{discussion}

Using a mean Milky Way star formation rate, we have demonstrated that feedback from supernovae and non-equilibrium photoionization from OB stars near the galactic midplane is not entirely sufficient to support and maintain the ionization state of the diffuse ionized gas. We find that the total mass at height is comparable to observations of total hydrogen density, with a slightly lower ionization state. This suggests that there could be a small missing ionizing component in the Milky Way, likely a combination of a population of hot evolved stars with the added ionization effects of cosmic rays.

Comparing to the simulations of \cite{kadofong}, our equivalent run assuming ionization equilibrium matches the vertical distribution from their MHD simulations, with neutral densities comparable to the Dickey-Lockman distribution, but less ionized gas than in the Reynolds layer. Our simulations with non-equilibrium ionization greatly increases the ionization state of the high altitude gas. The inclusion of non-equilibrium ionization is the crucial ingredient to produce widespread high altitude diffuse ionized gas in such simulations.

The time-dependent ionization calculations are particularly important for the DIG, as the timescale on which the ionizing flux available at height is varying is often shorter than the recombination time (tens of Myr) of the low density gas ($n<0.1{\rm cm}^{-3}$). The ionizing flux variability arises from the variability in the source ionizing luminosities with the rapid birth and death of massive stars, and also projection effects when dense midplane clumps move and can quickly block or open up photon paths to the diffuse high altitude gas.

For the low mass hot evolved stars in our simulations, we adopt a scale height of 700~pc and surface density of sources of $1000~\rm{kpc}^{-2}$. The ionizing luminosity and spectra of these low mass sources is uncertain. By exploring parameter space, we found that a population of hot evolved sources with a total luminosity of approximately $5\times10^{48} {\rm s}^{-1}\rm{kpc}^{-2}$ matches the observed DIG in many snapshots, while $1\times10^{49} {\rm s}^{-1}\rm{kpc}^{-2}$ has the effect of over-ionizing the high altitude gas, leaving less neutral gas at height. Although the hot evolved stars have low ionizing luminosities, they contribute to the ionized gas mass due to their quantity and scale height. Existing at heights of sometimes over 1~kpc above the midplane, these stars can ionize low density regions of the high altitude ISM, meaning only a few hot evolved stars in this environment can ionize a large volume of gas.

Our present simulations do not resolve individual molecular clouds or the very densest regions one might expect to find in the disc of the Milky Way. Because of this, we have introduced a parameter for the escape of ionising photons from molecular clouds. We find that in the fiducial run, an escape fraction of 0.1 replicates observed DIG densities most closely.

This escape fraction also aligns our escaping ionizing luminosity + HOLMES luminosity very closely with estimates of the total ionizing flux required to sustain the DIG. From \citet{digreview}, the ionising flux requirement is estimated as $2\times10^{6}~\rm{photons}~\rm{s}^{-1} \rm{cm}^{-2}$. Multiplying by two to account for both sides of the midplane, and by the $1~\rm{kpc}^2$ box area gives an estimate of $3.8\times10^{49}~\rm{photons}\, s^{-1}$ in our simulation box. The mean escaping ionizing flux plus HOLMES luminosity in our most realistic run (HOLMESLOW) is $3.6\times10^{49}~\rm{photons}~ s^{-1}$.

The required value of escape fraction is resolution dependent: increasing resolution would change the maximum densities achieved in the midplane, and also the porosity of the dense ISM. Future work in this area could include higher spatial resolution simulations with an adaptive grid along with explorations of the porosity of the ISM to ionizing photons.

The escape fraction comparison simulations of NOPHOTONS, DIM, fiducial and BRIGHT differ significantly in their evolution, with outflows being hindered in brighter simulations. This is due to the varying evolution of the SFR in these simulations. In this work (as in \citet{rathjen21}), self-consistency in the star formation algorithm has been prioritised over the direct investigation of the physical effect of the photoionization feedback. A more controlled investigation into the dynamical effect of ionizing photons with fixed SFRs represents future work.

There are some consequences of the size and shape of the simulation box. Periodic boundary conditions in the $x$ and $y$ axes means that we are in fact simulating a semi-infinite plane of a star forming galaxy. It is expected that this simplification is robust if gas and photons cannot interact with other gas and photons more than 1~kpc away. Another consequence of the 1~${\rm kpc}^{2}$ area box is the occasional asymmetry of the evolved structures at any one time (above-midplane structure sometimes being significantly different to below-midplane structure). This is likely a consequence of the thin simulation box, and a feature which one might not expect to observe in an entire galaxy simulation.

Our simulations do not include opacity due to interstellar dust grains. The work of \citet{kadofong} found that in similar simulations some 50\% of ionizing photons could be lost to absorption by dust. However, it was found that of those 50\%, most were lost within the most dense regions of the simulation. This could suggest that the effects of neglecting dust opacity in our simulations might be similar to overestimation of the escaping ionizing luminosity from the densest regions of the midplane. 

Another dust effect noted by \citet{barnes15} is H$\alpha$ emission from midplane H{\sc ii} regions being scattered by extraplanar dust. For some sightlines, dust scattered H$\alpha$ could comprise more than 20\% of the H$\alpha$ emission attributed to the DIG. With the $\rm{H}\alpha$ scaleheights in our simulations being toward the lower end of observed values, this dust scattering effect could provide extra $\rm{H}\alpha$ at height. In future work we will include the effects of dust on the absorption of ionizing photons and also dust scattering which may increase the scaleheight of H$\alpha$ emission.

Other similar work has included non-equilibrium ionization physics in radiation-hydrodynamcis simulations. The SILCC project (\citet{silcc,rathjen21,rathjen23}) has over time developed sophisticated treatments of various feedback mechanisms and non-equilibrium chemistry, and has investigated outflows and ISM properties as a function of disc surface density, and by turning different feedback mechanisms on and off to explore the effect of each mode of feedback. Similarly, in TIGRESS-NCR \citet{kim23} have presented state of the art simulations of the ISM under various galactic conditions, and carried out extensive analysis of the dynamics and phase state of their simulations. \citet{katz22} presents simulations coupling non-equilibrium metal chemistry to the RAMSES-RT radiation-hydrodynamics solver.

There are some physical processes present in these other non-equilibrium radiation-hydrodynamics simulations that are not included in our current simulations such as photoelectric heating from dust grains, magnetic fields, cosmic rays, sink particles and rotational shear. Our simulations therefore do not capture the accurate filling factors of warm and cold neutral gas (due to the absence of photoelectric heating), or pressure support by magnetic fields and cosmic rays present in the work of other groups. However, as supernovae provide the dominant feedback mechanism, we are in broad agreement with other groups as to the overall vertical extent and structure of neutral gas that is driven from the midplane to high altitude. We also do not use the sink particle prescription for star formation, and hence require a manual choice for the normalisation of the Kennicutt–Schmidt relation. While we employ a less self consistent algorithm for star formation, this allows us to set SFR as a simulation input rather than an output, giving us the ability to simulate a wide variety of ISM and SFR conditions. Another difference between our simulations and those of SILCC, TIGRESS-NCR and RAMSES-RT is the method of photon transport. With our MCRT photon shooting step, we possess the ability to work with more continuous source spectra, instead of a small number of photon energy bins. Whilst a subtle difference in the hydrogen only simulations, we expect this to be a key feature in the future development of the code to include the ionization structure of ISM metals, and ultimately produce accurate non-equilibrium emission line maps.

The focus of our current simulations is to study the conditions that can support and maintain the high altitude diffuse ionized gas, with particular emphasis on the role of non-equilibrium ionization effects in low density gas. While the details of the ISM structure in our simulations will certainly differ from those of other groups, we believe that in order to maintain high altitude DIG it is essential to include non-equilibrium ionization in simulations. 

While the focus of this work has been primarily on simulations of DIG within the Milky Way, there are extensive examples of extraplanar DIG in other galaxies (\citet{rand96,hoopes99,Jones2017,levy19}), often with comparable densities and scale heights to what is observed in the Milky Way. This opens avenues for further investigation into DIG properties as a function of other parameters e.g. surface density, star formation efficiency and HOLMES number density and luminosity. Furthermore, there is an abundance of vertically resolved emission line data from many of these edge on galaxies, with trends in emission line ratios similar to those seen in the Milky Way. These datasets provide an excellent opportunity to compare with the results of planned future extensions to the code investigating the non-equilibrium ionization state of ISM metals and resulting emission line intensities.

\section{Conclusions}
\label{conclusions}

In this work, we have further developed the simulations of \citet{bertkenny} to model the diffuse ionized gas in star forming regions of spiral galaxies. We have modified the simulations to track explicit photoionization heating terms and cooling rates from \citet{derijke}. This allowed us to implement supernovae feedback in the form of thermal and kinetic energy injection. To ensure the accuracy of the ionization balance calculation at high temperatures (now present due to supernova feedback), thermal collisional ionization of the gas was implemented to complement the existing photoionization schemes. A more complex and self-consistent algorithm for star formation was implemented, including a SFR variation with midplane density, and the ability to generate source positions within the densest regions of the midplane. A background type Ia supernova rate was also implemented.

The impact of a population of hot low mass evolved stars on the ionization state of the high altitude gas was investigated, as well as a study into the effect of varying star formation rates. The effect of varying SFR on the vertically resolved ISM is primarily seen in the shape of the neutral disc. Low values of SFR inhibits outflows and results in a sharp and dense disc, whereas high values tend to destroy the disc and power strong outflows to high altitudes. This is in broad agreement with other tall-box simulations of the ISM. \citet{rathjen23} investigated different surface densities in MHD simulations with self-consistent sink particle star formation, and found that powerful bursts of star formation would completely disrupt the disc, but low surface densities do not power outflows to support DIG-like structures. We similarly see close agreement to the simulations of \citet{kadofong} in our control simulations enforcing ionization equilibrium, whereby material is supported at height by the supernova feedback, but not held at a high enough ionization state to recreate the observed Reynolds layer.

We find that a star formation rate of $0.0032~ M_{\odot} \rm{yr}^{-1} \rm{kpc}^{-2}$ matches the Dickey-Lockman distribution of neutral gas. The required ionizing luminosity from hot evolved low mass stars to reproduce both the neutral and ionized densities at height is harder to constrain due to the stochasticity inherent to the simulations, and the dependence of this on other parameters. However, we find that a value of $5\times10^{48}~{\rm s}^{-1}\rm{kpc}^{-2}$ produces close to expected density of neutral and ionized material at high altitudes for many snapshots. This represents ~2.0\% of the total available ionizing luminosity. 

It is seen in all of our simulations that the volume filling factor of DIG rises with height above the midplane up to our simulation box ceiling at 3~kpc. This result can be used to reconcile tensions between H$\alpha$ and pulsar dispersion measurements of DIG scale height, with the H$\alpha$ derived electron scale height often presenting as less than 75\% of the true value.

The most important code development relating to the simulation of low density diffuse ionized gas is the inclusion of a time-dependent ionization state calculator to follow the ionization and recombination of the DIG. If ionization equilibrium is assumed, a Milky Way-like DIG layer is rarely formed. At DIG densities the recombination timescales can exceed tens of millions of years. We find that the calculation of the ionization state in a time-dependent manner is crucial for the study of ionized gas at DIG densities.

In future work we plan to explore the effects of other physical processes on the DIG structure and also produce synthetic emission lines from other elements (He, C, N, O, Ne, S) by extending our non-equilibrium ionization scheme to include those elements also. The planned extension of our non-equilibrium ionization scheme will then allow a more detailed comparison of our simulations with DIG emission line observations.


\section*{Acknowledgements}

We thank Steffi Walch-Gassner and Tim-Eric Rathjen for many productive discussions relating to the development of our simulations. LM acknowledges financial support from a UK-STFC PhD studentship. CP was funded by a UK-STFC PRDA during the early development of this work. DK acknowledges support from an HST Archival Theory Program 17060, provided by NASA through a grant from the Space Telescope Science Institute, which is operated by the Association of Universities for Research in Astronomy, Inc., under NASA contract NAS5-26555.

\section*{Data Availability}

The simulation data in this work will be shared on reasonable request.



\bibliographystyle{mnras}
\bibliography{biblio} 








\bsp	
\label{lastpage}
\end{document}